\documentclass[a4paper,fleqn,usenatbib,natbib]{mnras}
\usepackage{physics}
\usepackage{xspace}
\usepackage{epsfig}
\usepackage{amssymb}
\usepackage{amsmath}
\usepackage{newtxtext,newtxmath}
\usepackage{xcolor}
\usepackage{slashbox}
\usepackage{array}
\usepackage{multirow}
\usepackage{mathtools}

\newcommand{\be}{\begin{equation}}
\newcommand{\e}{\end{equation}}
\newcommand{\bear}{\begin{eqnarray}}
\newcommand{\ear}{\end{eqnarray}}

\newcommand{\comment}[1]{}

\def\xi{x^{i}_{\ion{H}{i}}\,}
\def\xh1{x_{\ion{H}{i}\,}}
\def\xb{\bar{x}_{\ion{H}{i}}}
\def\Ph1{P_{\ion{H}{i}}}
\def\Bh1{B_{\ion{H}{i}}}
\def\eh1{\eta_{\ion{H}{i}}}

\def\HI{\ion{H}{i}\xspace}
\def\HII{\ion{H}{ii}\xspace}
\def\P{\hat{P}}

\def\mp{\, {\rm Mpc}^{-1}}

\begin{document}

\title[RSD in EoR 21-cm bispectrum]{Redshifted 21-cm Bispectrum I: Impact of the Redshift Space Distortions on the Signal from the Epoch of Reionization}
\author[Majumdar et al.]{Suman Majumdar$^{1,2}$\thanks{suman.majumdar@iiti.ac.in},
Mohd Kamran$^{1}$, Jonathan R. Pritchard$^{2}$,
\newauthor Rajesh Mondal$^{3}$, Arindam Mazumdar$^{4}$, Somnath Bharadwaj$^{4}$, Garrelt Mellema$^{3}$
\\
$^{1}$Discipline of Astronomy, Astrophysics and Space Engineering, Indian Institute of Technology Indore, Simrol, Indore 453552, India\\
$^{2}$Department of Physics, Blackett Laboratory, Imperial College, London SW7 2AZ, U. K.\\
$^{3}$Department of Astronomy and Oskar Klein Centre, Stockholm University, AlbaNova, SE-10691 Stockholm, Sweden\\
$^{4}$Department of Physics and Centre for Theoretical Studies, Indian Institute of Technology Kharagpur, Kharagpur - 721 302, India\\
}
\date{Accepted 2020 October 08. Received 2020 October 08; in original form 2020 July 12}

\maketitle

\begin{abstract}
The bispectrum can quantify the non-Gussianity present in the redshifted 21-cm signal produced by the neutral hydrogen ($\HI$) during the epoch of reionization (EoR). Motivated by this, we perform a comprehensive study of the EoR 21-cm bispectrum using simulated signals. Given a model of reionization, we demonstrate the behaviour of the bispectrum for all unique triangles in $k$ space. For ease of identification of the unique triangles we parametrize the $k$-triangle space with two parameters, namely the ratio of the two arms of the triangle ($n=k_2/k_1$) and the cosine of the angle between them ($\cos{\theta}$). Furthermore, for the first time we quantify the impact of the redshift space distortions (RSD) on the spherically averaged EoR 21-cm bispectrum in the entire unique triangle space. We find that the real space signal bispectra for small and intermediate $k_1$-triangles ($k_1 \leq 0.6 \,\mp$) is negative in most of the unique triangle space. It takes a positive sign for squeezed, stretched and linear $k_1$-triangles, specifically for large $k_1$ values ($k_1 \geq 0.6 \,\mp$). The RSD affects both the sign and magnitude of the bispectra significantly. It changes (increases/decreases) the magnitude of the bispectra by $50-100\%$ without changing its sign (mostly) during the entire period of the EoR for small and intermediate $k_1$-triangles. For larger $k_1$-triangles, RSD affects the magnitude by $100-200\%$ and also flips the sign from negative to positive. We conclude that it is important to take into account the impact of RSD for a correct interpretation of the EoR 21-cm bispectra.
\end{abstract}


\section{Introduction}
\label{sec:intro}

After the Big Bang the Universe gradually cooled down and once it reached a temperature of about 3000~K most of the hydrogen in the Universe went through a phase change from ionized (\HII) to neutral (\HI) during the so-called epoch of recombination. After this it stayed neutral until the first sources of light formed. These sources are thought to have generated enough radiation in the X-ray and ionizing UV bands (see e.g. \citealt{barkana09, dayal18} for properties of the sources of reionization) to gradually heat and ``re''-ionize most of the neutral hydrogen (see e.g. \citealt{furlanetto06, pritchard08, pritchard12, choudhury09} for reviews). This final phase transition of hydrogen marks one of the least understood periods in the history of our universe: the Epoch of Reionization (EoR). Only a few indirect observations guide our present understanding of this epoch. These are the cosmic microwave background radiation (CMB) \citep[see e.g.][] {komatsu11,planck16}, the absorption spectra of high redshift quasars \citep[see e.g.][] {becker01, fan03, white03, barnett17} and the luminosity function and clustering properties of Lyman-$\alpha$ emitters \citep[see e.g.][]{trenti10, ouchi10, jensen13b, choudhury14, bouwens16, ota17}. They jointly suggest that this phase transition of hydrogen spans a wide redshift range, $6 \lesssim z \lesssim 15$ \citep[see e.g.][]{alvarez06, mitra15, robertson15, bouwens15}. However, these observations do not provide the precise duration and timing of reionization and they do not put strong constraints either on the properties of the main sources of ionization and heating nor on the typical size distribution of the ionized regions at different stages of reionization.

The redshifted 21-cm line, originating from spin-flip transitions in $\HI$ atoms, is the most promising tool for the direct observation of the EoR and can potentially answer many of its fundamental puzzles, as mentioned above. The brightness temperature of the redshifted 21-cm line directly probes the \HI density at the epoch where the radiation originated. One can, in principle, track the evolution of \HI during the entire reionization period by observing this line at different redshifts.

Motivated by this prospect a number of low frequency radio interferometers, such as the GMRT \citep{paciga13}, LOFAR \citep{mertens20}, MWA \citep{barry19, li19}, PAPER \citep{kolopanis19} and 21CMA \citep{wang2013} are competing to achieve the first statistical detection of the redshifted 21-cm signal from the EoR. However, as yet none of these has produced a successful detection of the signal, largely due to the complications of separating the signal from the $\sim\!4 -5$ orders of magnitude stronger foreground emission \citep[e.g.][]{dimatteo02, ali08, jelic08, ghosh12}, and system noise \citep{morales05, mcquinn06}. Only weak upper limits on the expected 21-cm signal have been obtained \citep{paciga13, mertens20, barry19, li19, kolopanis19, 2020MNRAS.493.4711T}.

Once an optimal method of separating the signal from the foreground contaminated data is achieved, the first generation interferometers will probably detect the signal via statistical quantities such as the variance \citep[e.g.][]{patil14, watkinson14, watkinson15}, the multi-frequency angular power spectrum \citep[e.g.][] {datta07a, mondal18, mondal19, mondal20a}) and the power spectrum \citep[e.g.][] {pober14, patil17}, as these lead to the high signal-to-noise ratios (SNR). The power spectrum has been shown to probe many important features of the signal \citep[see e.g.][]{bharadwaj04, barkana05, datta07a, mesinger07, lidz08, choudhury09b, mao12, majumdar16b, jensen13} and thus can be used for the EoR parameter estimation \citep{greig15,greig19,koopmans15, ghara20, mondal20}.

However, only for a Gaussian random field does the power spectrum provide a complete statistical description. The fluctuations in the EoR 21-cm signal are dictated by the interplay between the underlying matter density and the evolving distributions of the ionized regions{\footnote{When the spin temperature $T_{\rm S}$ has saturated over the CMB temperature $T_{\rm CMB}$ i.e. $T_{\rm S} \gg T_{\rm CMB}$.}}. These make the signal highly non-Gaussian. The power spectrum is incapable of capturing this non-Gaussianity in the signal, however the error in the power spectrum estimation (cosmic covariance) will be significantly affected by it \citep{mondal15,mondal16,mondal17}. The position dependent power spectrum, estimated by dividing a large survey volume into several smaller sub-volumes and then calculating the power spectra of those sub-volumes, can however quantify to some degree the signal correlation (mode coupling) between small and large length scales which is caused by the signal's non-Gaussianity \citep{giri19}.

Quantifying the non-Gaussianity of the signal requires the use of higher order statistics. One-point statistics such as the skewness and kurtosis provide a straightforward means to achieve this \citep[see e.g.][]{harker09, watkinson14, watkinson15, shimabukuro15a, kubota16}. They capture the general level of non-Gaussianity integrated over the range of scales from which they are measured. However, as one-point statistics, they are incapable of quantifying the correlation of the signal between different length scales.

The bispectrum, which is estimated through the product of the Fourier transform of the signal for a set of three wave numbers (${\bf k}$) that form a closed triangle in Fourier space, is on the other hand capable of quantifying the correlations of the signal between different Fourier modes. It is apparent that a successful detection of the signal bispectrum, the Fourier equivalent of the three point correlation function, will require more sensitivity than needed for the signal power spectrum. \citet{trott19} tried to put an upper limit on the signal bispectrum using the observations with the MWA Phase II array. Measurement of the bispectrum not only characterise the non-Gaussianity of the signal but will also constitute an important  confirmative detection of the EoR 21-cm signal, as any claimed measurement of the power spectrum could contain contributions from residual foregrounds or noise.

Understanding the characteristics of the EoR 21-cm bispectrum is more relevant now in view of the construction of the more sensitive next generation radio interferometers HERA  \citep{pober14,ewallwice14} and the SKA1-LOW \citep{koopmans15}. SKA1-LOW is expected to see first light in around 2026. The detection and characterization of EoR 21-cm signal is one of its key science goals. As argued above, a measurement of the 21-cm bispectrum should be an integral part of this both as a confirmation of any claimed power spectrum measurement and as a quantification of the non-Gaussianity of the signal.

Theoretical efforts to characterize the EoR 21-cm bispectrum started with analytical models \citep{bharadwaj05a, ali06} which were followed by more detailed radiative transfer and semi-numerical simulations of the signal \citep{yoshiura14, shimabukuro16, majumdar18, watkinson19, hutter19, saxena20}. In a previous paper \citep[][, hereafter Paper I]{majumdar18}, we quantified the EoR 21-cm bispectrum using an ensemble of semi-numerically simulated 21-cm signals and for a variety of $k$-triangles (e.g.\  equilateral, isosceles, etc.). Through an analytical model for the 21-cm signal fluctuations, we showed that there are two competing components of the signal driving the non-Gaussianity in the signal: fluctuations in the neutral fraction ($\xh1$) field and fluctuations in the matter density field. We further showed that the sign of the bispectrum works as a unique marker to identify which of these two components is driving the non-Gaussianity: the bispectrum will have a negative sign when the non-Gaussianity is driven by the distribution of the ionized regions and it will be positive when the non-Gaussianity is driven by the matter density fluctuations. We also proposed that this sign change in the bispectrum when viewed as a function of triangle configuration and reionization history can be used as a confirmative test for the detection of the EoR 21-cm signal. \citet{hutter19}, using a set of semi-numerical simulations of the EoR, independently arrived at similar conclusions. \citet{saxena20} have studied the impact of different dark matter models on the EoR 21-cm bispectrum and have shown that the differences in the signal bispectrum is more prominent for different dark matter models compared to the differences in the signal power spectrum. 
Analysing a set of radiative transfer simulations of the Cosmic Dawn (CD), when X-ray heating played a crucial role in determining the brightness temperature fluctuations, \citet{watkinson19} showed that amplitude and sign of the CD 21-cm bispectrum depends on the distribution and size of the heated regions. In a follow up work, \citet{watkinson20} investigated how the signal bispectrum is affected by the presence of foreground signals and whether foreground mitigation through subtraction or avoidance would be better for detecting it. It should be pointed out that due to the specific definition of the bispectrum estimator used by \citet{yoshiura14, shimabukuro16}, these authors were unable to capture the sign of the bispectrum, which \citet{majumdar18}, \citet{hutter19} and \citet{watkinson19} found to be an important feature of this statistic.

The coherent inflows and outflows of matter into overdense and underdense regions respectively, produce an additional red-- or blueshift in the 21-cm signal on top of the cosmological redshift, changing the contrast of the 21-cm signal, and making it anisotropic along the LoS. This apparent LoS anisotropy in the signal is known as redshift space distortions (RSD) and was first highlighted by \citet{bharadwaj04, bharadwaj05} and \citet{ali05} in the context of the 21-cm signal from the CD/EoR. Using analytical models for the signal they showed that the peculiar velocities will significantly change the amplitude and the shape of the 21-cm power spectrum. Their analytical predictions were later independently tested and validated through both radiative transfer and semi-numerical simulations by \citet{mao12, majumdar13, majumdar14, majumdar16, jensen13, ghara15a,  fialkov15}. All of these studies independently report a significant change in the shape and amplitude of signal power spectrum, sometimes by a factor of $\sim 3$ or more depending on the $k$ mode and stage of reionization. This implies that if the effect of RSD is not taken into account, it may lead to an incorrect interpretation of the signal.

All of the previous CD/EoR 21-cm bispectrum studies have been performed for the real space signal i.e.\ without taking into account redshift space distortions. In this article we aim to quantify the impact of RSD on the shape, amplitude and sign of the EoR 21-cm signal bispectrum at different stages of reionization using an ensemble of simulated 21-cm signals. Additionally, we present the first comprehensive view of the signal's non-Gaussianity by calculating the bispectrum for all possible unique $k$-triangles in Fourier space. The earlier EoR 21-cm bispectrum studies mentioned above only considered a few specific $k$-triangle configurations when estimating the bispectrum. We first identify all possible unique $k$-triangles in terms of the triangle parameters, following the formalism of \citet{bharadwaj20} and then estimate the spherically averaged real and redshift space bispectra to quantify the impact of the redshift space distortions. Lastly, we provide a physical interpretation of our results based on the quasi-linear model of brightness temperature fluctuations proposed by \citet{mao12}.

The structure of this paper is as follows. In Section \ref{sec:bispec_th}, we briefly describe the algorithm that we adopt to estimate the bispectrum from the simulated signal. We also define the unique triangle configurations (Section \ref{sec:bispec_uni_tri}) that we consider for our bispectra estimation. Section \ref{sec:sim} briefly describes our method to generate mock 21-cm data sets. In Section \ref{sec:results}, we discuss and interpret our estimated bispectra for all unique triangle configurations as well as a quasi-linear model to understand the results (Section \ref{subsec:quasi}). Finally, in Section \ref{sec:summary} we summarise our findings.

Throughout this paper, we have used the Planck+WP best fit values of
cosmological parameters $h = 0.6704$, $\Omega_{\mathrm{m}} = 0.3183$,
$\Omega_{\Lambda} = 0.6817$, $\Omega_{\mathrm{b}} h^2 = 0.022032$,
$\sigma_8 = 0.8347$ and $n_s = 0.9619$ \citep{planck14}.

\section{Bispectrum estimation for all unique triangle configurations}
\label{sec:bispec_th}

\subsection{Bispectrum estimator for the simulated 21-cm signal}
\label{sec:bispec_est}

\begin{figure*}
  \includegraphics[width=0.5\textwidth,angle=0]{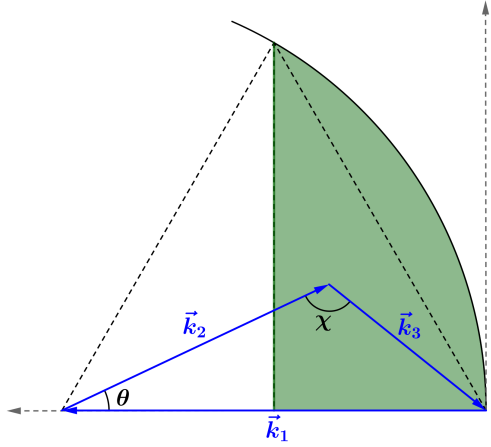}
  \includegraphics[width=0.47\textwidth,angle=0]{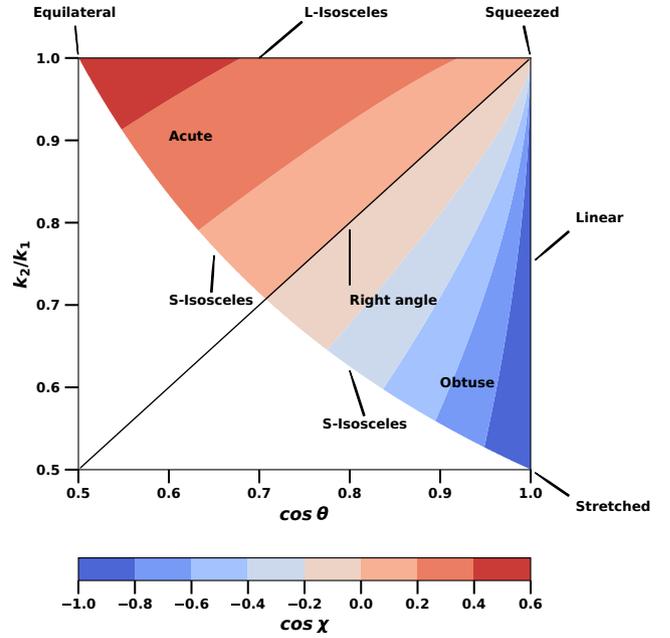}
  \caption{The left panel shows the definition of unique triangles in ${\bf k}$-space. The right panel shows the variation of $\cos{\chi}$ with $k_2/k_1$ and $\cos{\theta}$ in unique triangles. The unique triangle parameter space is defined following \citet{bharadwaj20}.}
  \label{fig:triangle_uni}
\end{figure*}

The bispectrum $B_{{\rm b}}({\bf k}_1, {\bf k}_2, {\bf k}_3)$ of the 21-cm brightness temperature fluctuations $ \delta T_{{\rm b}}({\bf x})$ can be defined as 
\begin{equation}
    \langle \Delta_{{\rm b}}({\bf k}_1) \Delta_{{\rm b}}({\bf k}_2)
 \Delta_{{\rm b}}({\bf k}_3)\rangle = V \delta^{{\rm K}}_{{\bf k}_1 +{\bf k}_2 +
{\bf k}_3, 0}\, B_{{\rm b}}({\bf k}_1, {\bf k}_2, {\bf k}_3) \, ,
\end{equation}
where $\Delta_{{\rm b}}({\bf k})$ is the Fourier transform of $\delta T_{{\rm b}}({\bf x})$ and $\delta^{{\rm K}}_{{\bf k}_1 +{\bf k}_2 +{\bf k}_3, 0}$ is the Kronecker delta function, equal to $1$ when ${\bf k}_1 +{\bf k}_2 +{\bf k}_3 = 0$ and $0$ otherwise. The $\delta^{{\rm K}}_{{\bf k}_1 +{\bf k}_2 +{\bf k}_3, 0}$ ensures that only those ${\bf k}$ triplets contribute to the bispectrum which form a closed triangle (see left panel of Figure \ref{fig:triangle_uni}). The angular brackets denote ensemble average of the target statistic. For brevity we drop the subscript ``b'' when describing the brightness temperature from this point on-wards.

The estimator, that one can use to compute the bispectrum from the observed or simulated data, can be defined for the $m^{\rm th}$ triangle configuration bin as
\begin{equation}
  \hat{B}_{m}({\bf k}_1, {\bf k}_2, {\bf k}_3) = \frac{1}{N_{{\rm tri}}V} \sum_{
[{\bf k}_1 +{\bf k}_2 +{\bf k}_3 =0] \in m }\Delta ({\bf k}_1) \Delta ({\bf k}_2
) \Delta ({\bf k}_3) \, ,
\label{eq:bispec_est}
\end{equation}
where $N_{{\rm tri}}$ is the number of closed triangles contributing to the $m^{\rm th}$ triangle bin for which one estimates the bispectrum. As discussed in detail in section~2 of Paper I, the bispectrum of a real field such as the 21-cm signal, also is a real quantity.

To estimate the bispectra from the simulated signal cubes we adopt the method described in Paper I. This method is a direct implementation of Equation \eqref{eq:bispec_est} along with the following two equations of constraints:  
\begin{equation}
k_2/k_1 = n\,,
\label{eq:ratio}
\end{equation}
and 
\begin{equation}
\frac{{\bf k}_1 . {\bf k}_2}{k_1 k_2} = -\cos\theta\, .
\label{eq:theta}
\end{equation}
The angle $\theta$ (between ${\bf k_1}$ and ${\bf k_2}$) is defined in Figure \ref{fig:triangle_uni}. Using the above equations one can parametrize the triangle configurations in terms of the values of $k_1$, $n$ and $\cos\theta$, which jointly determine the size and shape of the triangle. Another way of interpreting the shape of the triangle is through the angle $\chi$, the angle between arms $k_2$ and $k_3$ of the $k$-triangle, which is dependent on $n$ and $\cos{\theta}$ via 
\begin{equation}
    \cos{\chi} = \frac{n^2+[1+n^2-2n\cos{\theta}]-1}{2n \sqrt{1+n^2-2n\cos{\theta}}}\,.
    \label{eq:cos_chi}
\end{equation}

The computationally intensive part of the bispectrum estimation algorithm is the search for closed ${\bf k}$-triangles in a gridded Fourier space. Equations \eqref{eq:ratio} and \eqref{eq:theta} effectively eliminate two nested {\it for} loops from the triangle search algorithm. This makes the calculation of bispectra from gridded data much more efficient. For a more detailed discussion of this method of bispectrum estimation we refer the reader to section 2 of Paper I. 

\subsection{Unique $k_1$-triangle configurations}
\label{sec:bispec_uni_tri}

In Paper I, our analysis of 21-cm bispectrum was limited to only a few types of ${\bf k}$-triangles, namely equilateral, isosceles and triangles with $n = 2, \, 5, \, {\rm and} \, 10 $. A comprehensive view of the signal non-Gaussianity as captured by the bispectrum requires the calculation of the bispectrum for all possible ${\bf k}$-triangles. However, not all ${\bf k}$-triangles will be unique in their shape and size. To identify unique ${\bf k}$-triangles in the Fourier space we follow the definition of \citet{bharadwaj20} and impose the following additional conditions:
\begin{eqnarray}
&&{\bf k}_1 \geq {\bf k}_2 \geq {\bf k}_3
\label{eq:k_relation}\\
&&0.5 \leq \cos\theta \leq 1.0
\label{eq:costheta}\\
&&0.5 \leq n \leq 1.0\, .
\label{eq:n}
\end{eqnarray}
The triangles that satisfy these conditions are confined to the region of the $n-\cos\theta $ parameter space where $n \cos\theta \geq 0.5$. This effectively means that the location of the tip which connects ${\bf k_2}$ and ${\bf k_3}$, as we vary $\cos{\theta}$ and $n$, will be restricted to the shaded region shown in the left panel of Fig.~\ref{fig:triangle_uni}. Any triangle that falls outside of this region, can be transformed into one that does by relabelling its sides. Note that Equation \eqref{eq:k_relation} ensures that in this parameterization of unique $k$-triangles, ${\bf k_1}$ remains the largest side under all circumstances. This point is important as we bin our bispectra estimates based on the value of $k_1$. 

For ease of identification of $k$-triangles of different shapes in  $n-\cos{\theta}$ parameter space we indicate them in the right panel of Fig.~\ref{fig:triangle_uni} and also list them here:
\begin{itemize}
    \item L-isosceles are triangles with $n = 1.0$ and $0.5 \lesssim \cos{\theta} \lesssim 1.0$ i.e. $0 \lesssim \cos{\chi} \lesssim 0.5$. These triangles have  $k_1=k_2$.
    \item The $n\cos{\theta} = 0.5$ arc in the parameter space defines S-isosceles triangles. These triangles have $k_2 = k_3$.
    \item The junction of the L and S isosceles triangles represents equilateral triangle i.e. $\cos{\theta}=0.5$ and $n=1.0$ (i.e. $\cos{\chi}=0.5$).
    \item Triangles with $\cos{\theta}\rightarrow1.0\,$ and $0.5 \lesssim n \lesssim 1.0$ are linear triangles. At these limits all three $k$s become collinear (i.e. $\cos{\chi}\rightarrow-1.0$).
    \item Triangles with $\cos{\theta}=n$ are right angle triangles (i.e. $\cos{\chi}=0$).
    \item The junction of L-isosceles, linear and right angle triangles represents squeezed triangles where $\cos{\theta}=n=1.0$ (i.e. $\cos\chi=0$). For squeezed triangles the smallest arm $k_3 \rightarrow 0$.
    \item  The junction of S-isosceles and linear triangles defines stretched triangles where $\cos\theta=1.0$ and $n=0.5$ (i.e. $\cos{\chi}=-1.0$).
    \item Triangles with $\cos{\theta}<n$ are acute angle triangles (i.e. $\cos{\chi>0}$).
    \item Triangles with $\cos{\theta}>n$ are obtuse angle triangles (i.e. $\cos{\chi<0}$).
\end{itemize}

To estimate bispectra from the simulated signal cubes we have divided the region $n\cos{\theta} \geq 0.5$ in the triangle parameter space into a grid of resolution $\Delta{\cos{\theta}} = 0.01$, $\Delta{n} = 0.05$. Additionally, we have divided the entire $k_1$ range (defined by $k_\mathrm{min} = 2\pi/$[box size]  and $k_\mathrm{max} = 2\pi/2$[grid spacing]) into $15$ logarithmic bins. For our simulation data (see Section~\ref{sec:sim}) $k_\mathrm{min} = 0.03$  Mpc$^{-1}$ and $k_\mathrm{max} = 5.61$ Mpc$^{-1}$. The estimated bispectra are averaged over these $k_1$ bins.

\section{Quasi-linear model for the redshift space EoR 21-cm bispectrum}
\label{subsec:quasi}
The main aim of this article is to quantify the impact of the redshift space distortions on the EoR 21-cm bispectrum, an effect which has not been considered in any of the previous studies.  However, it would be easier to interpret this impact if we can analyze it with the help of an analytical model \citep[e.g.][] {bharadwaj05, lidz08, mao12}. In this section we use the prescription of \citet{mao12} to construct such a model for the redshift space 21-cm bispectrum. 

The redshift space bispectrum depends on how the three vectors ${\bf k_1}$, ${\bf k_2}$ and ${\bf k_3}$, that form a closed triangle in the Fourier space, are oriented with respect to the LoS. We use $\mu_1,\mu_2$ and $\mu_3$ respectively to denote the cosine of the angles that ${\bf k_1}$, ${\bf k_2}$ and ${\bf k_3}$ make with the LoS. However, as pointed out by \citet{bharadwaj20},  $\mu_1, \mu_2$ and $\mu_3$ are not independent of each other, in fact they refer to a particular triangle whose shape and size are fixed for a fixed set of values of $k_1$, $n$ and $\cos{\theta}$. These authors also showed that all possible orientations of a triangle of fixed shape and size with respect to the LoS can be obtained by performing rigid body rotations of the triangle (see sections 2 and 3 of \citealt{bharadwaj20} for more details). This is a crucial point that should be kept in mind while interpreting the redshift space spherically averaged signal bispectrum. 

Adopting the quasi-linear model of \citet{mao12}, the redshift space 21-cm bispectrum for a fixed ${\bf k}$-triangle can be written as the sum of different auto and cross bispectra between two fields, namely the total hydrogen density (${\rho_{\rm H}}$) and the neutral hydrogen density (${\rho_{\HI}}$):  
\begin{multline}
     B^{\rm {s, qlin}}({\bf k_1},{\bf k_2},{\bf k_3}) = \Big(\widehat{\delta T_{\rm b}}\Big)^3 \Big[B^r_{\Delta_{\rho_{\HI}}, \Delta_{\rho_{\HI}}, \Delta_{\rho_{\HI}}}+ \\ \Big(
     \mu_3^2 B^r_{\Delta_{\rho_{\HI}}, \Delta_{\rho_{\HI}}, \Delta_{\rho_{\rm H}}}+
     \mu_2^2 B^r_{\Delta_{\rho_{\HI}}, \Delta_{\rho_{\rm H}}, \Delta_{\rho_{\HI}}}+
     \mu_1^2 B^r_{\Delta_{\rho_{\rm H}}, \Delta_{\rho_{\HI}}, \Delta_{\rho_{\HI}}} \Big) \\+ \Big(
     \mu_1^2 \mu_2^2 B^r_{\Delta_{\rho_{\rm H}}, \Delta_{\rho_{\rm H}}, \Delta_{\rho_{\HI}}}+
     \mu_1^2 \mu_3^2 B^r_{\Delta_{\rho_{\rm H}}, \Delta_{\rho_{\HI}}, \Delta_{\rho_{\rm H}}}+
     \mu_2^2 \mu_3^2 B^r_{\Delta_{\rho_{\HI}}, \Delta_{\rho_{\rm H}}, \Delta_{\rho_{\rm H}}} \Big)\\ +
     \mu_1^2 \mu_2^2 \mu_3^2 B^r_{\Delta_{\rho_{\rm H}}, \Delta_{\rho_{\rm H}}, \Delta_{\rho_{\rm H}}} \Big]
     \label{eq:B_qlin}
\end{multline}
where
\begin{equation}
    \widehat{\delta T_{\rm b}}(z_{\rm cos}) = 27  \bar{x}_{\HI}(z_{\rm cos})  \Bigg( \frac{\Omega_{\rm b} h^2}{0.023} \Bigg) \Bigg( \frac{0.15}{\Omega_{\rm m} h^2} \frac{1+z_{\rm cos}}{10} \Bigg)^{1/2} {\rm mK} \, ,
    \label{eq:tb_average}
\end{equation}
and the superscripts $s$ and $r$ represent terms in redshift and real space, respectively. It is convenient to represent the anisotropy in the signal bispectrum by decomposing it in the orthonormal basis of spherical harmonics ${Y}^m_{\ell}$. The different  angular multipole moments of the RSD bispectrum thus can be expressed as:
\begin{multline}
     \bar{B}^m_{\ell}(k_1, n, \cos{\theta}) = \Big(\widehat{\delta T_{\rm b}}\Big)^3 \Big[
    \delta_{\ell,0}   B^r_{\Delta_{\rho_{\HI}}, \Delta_{\rho_{\HI}}, \Delta_{\rho_{\HI}}} \\+ \Big(
     [\overline{\mu_3^2}]^m_{\ell}  B^r_{\Delta_{\rho_{\HI}}, \Delta_{\rho_{\HI}}, \Delta_{\rho_{\rm H}}}+
     [\overline{\mu_2^2}]^m_{\ell}   B^r_{\Delta_{\rho_{\HI}}, \Delta_{\rho_{\rm H}}, \Delta_{\rho_{\HI}}} \\+
    [\overline{\mu_1^2}]^m_{\ell}      B^r_{\Delta_{\rho_{\rm H}}, \Delta_{\rho_{\HI}}, \Delta_{\rho_{\HI}}} \Big)\\ + \Big(
    [\overline{\mu_1^2 \mu_2^2}]^m_{\ell}   B^r_{\Delta_{\rho_{\rm H}}, \Delta_{\rho_{\rm H}}, \Delta_{\rho_{\HI}}}+
     [\overline{\mu_1^2 \mu_3^2}]^m_{\ell}   B^r_{\Delta_{\rho_{\rm H}}, \Delta_{\rho_{\HI}}, \Delta_{\rho_{\rm H}}} \\+
     [\overline{\mu_2^2 \mu_3^2}]^m_{\ell}  B^r_{\Delta_{\rho_{\HI}}, \Delta_{\rho_{\rm H}}, \Delta_{\rho_{\rm H}}} \Big)\\ +    
     [\overline{\mu_1^2 \mu_2^2 \mu_3^2}]^m_{\ell} B^r_{\Delta_{\rho_{\rm H}}, \Delta_{\rho_{\rm H}}, \Delta_{\rho_{\rm H}}} \Big]\,.
     \label{eq:B_qlin1}
\end{multline}
The $\bar{B}^m_{\ell}$ here represents the value of a specific multipole averaged over all possible orientations of a fixed triangle. In this paper we are interested in the spherically averaged redshift space signal bispectrum, which is nothing but the monopole moment of the Equation (\ref{eq:B_qlin}), i.e. Equation (\ref{eq:B_qlin1}) for $m = 0$ and $\ell = 0$. For the monopole moment{\footnote{To quantify the impact of the RSD precisely one in principle would need to estimate all non-zero angular multipole moments of Equation \eqref{eq:B_qlin}, the direction dependent bispectrum \citep[see e.g.][]{mazumdar20}. However, our aim here is to quantify the impact of RSD on the spherically averaged bispectrum. This is why we concentrate only on the monopole moment of the RSD bispectrum in this paper.}}, different coefficients in Equation (\ref{eq:B_qlin1}) will take the form
\begin{equation}
\delta_{0,0} = 1\,,
\label{eq:B_mu0}
\end{equation}
\begin{equation}
[\overline{\mu_1^2}]_0^0 = [\overline{\mu_2^2}]_0^0 = [\overline{\mu_3^2}]_0^0 = \frac{1}{3}\,,
\label{eq:B_m2}
\end{equation}
\begin{equation}
[\overline{\mu_1^2 \mu_2^2}]_0^0 =\frac{1}{15} \left(2 \cos^2{\theta}+1\right)\,,
\label{eq:B_mu4_1}
\end{equation}
\begin{equation}
[\overline{\mu_2^2 \mu_3^2}]_0^0 =\frac{2 \cos^2{\theta}+3 n^2-6n \cos{\theta} +1}{15 s^2}\,,
\label{eq:B_mu4_2}
\end{equation}
\begin{equation}
[\overline{\mu_3^2 \mu_1^2}]_0^0 =\frac{\left(2 \cos^2{\theta} +1\right) n^2-6n \cos{\theta}  
+3}{15 s^2}\,,
\label{eq:B_mu4_3}
\end{equation}
\begin{equation}
[\overline{\mu_1^2 \mu_2^2 \mu_3^2}]_0^0 =\frac{4\left(n^2+1\right) \cos^2{\theta}+n^2-4n \cos^3{\theta} -6n \cos{\theta}  +1}{35 s^2}\,
\label{eq:B_mu6}
\end{equation}
where $s = \sqrt{1-2n \cos{\theta} +n^2}$. 

The above equations demonstrate that out of the eight coefficients of the monopole moment, four are dependent on the shape of the triangle (Equations \eqref{eq:B_mu4_1}- \eqref{eq:B_mu6}). These shape dependent coefficients vary in the range\footnote{The detailed $k$-triangle shape dependence of these four coefficients are shown in Figure \ref{fig:monopole} in the Appendix \ref{sec:app1}.} $0.01-0.20$. Equation \eqref{eq:B_qlin1} also shows that in the absence of any redshift space distortions, the observed spherically averaged 21-cm bispectrum i.e. the monopole of Equation \eqref{eq:B_qlin1} will reduce to $\Big(\widehat{\delta T_{\rm b}}\Big)^3 B^r_{\Delta_{\rho_{\HI}}\Delta_{\rho_{\HI}}, \Delta_{\rho_{\HI}}}$ (first term in the R.H.S.). We thus identify all R.H.S. terms apart from the first one as the redshift space correction (RC) terms to the real space spherically averaged bispectrum ($m=0,\,{\rm and}\, \ell=0$). Among the seven RC terms we label the sum of the first three terms as $B_{\mu^2-{\rm RC}}$, the sum of the next three as $B_{\mu^4-{\rm RC}}$ and the last as $B_{\mu^6-{\rm RC}}$. We will use Equation \eqref{eq:B_qlin1} together with these notations as a tool to provide some physical interpretation of the simulated redshift space 21-cm bispectra below in Section~\ref{sec:interpretation}.
\section{Simulating the redshifted 21-cm signal from the EoR}
\label{sec:sim}
For our study we use the redshifted EoR 21-cm brightness temperature maps from the simulations of \citet{mondal17} at the seven different redshifts 13, 11, 10, 9, 8, 7.5 and 7. In this section we briefly summarize the semi-numerical technique used to generate the redshifted signal. A detailed description of these simulations can be found in \citet{mondal17, mondal18}. Our simulation method for the 21-cm signal is divided into three main steps. In the first step, we generate dark matter distributions at the desired redshifts using a publicly available parallelized particle mesh (PM) \textit{N}-body code\footnote{\url{https://github.com/rajeshmondal18/N-body}}. In the second step, we identify collapsed dark matter halos using a publicly available Friends-of-Friend (FoF) code\footnote{\url{https://github.com/rajeshmondal18/FoF-Halo-finder}}. The final step is to generate ionization fields using a publicly available semi-numerical reionization code\footnote{\url{https://github.com/rajeshmondal18/ReionYuga}} which is based on the excursion set formalism of \citet{furlanetto04b}. The third step closely follows the inside-out reionization model of \citet{choudhury09b}. Here the assumptions are that the hydrogen follows the underlying matter distribution and that luminous sources form within the halos. Finally, the resulting neutral hydrogen fields are mapped to redshift space to generate the redshifted 21-cm signal following \citet{majumdar13}. 

The $N$-body simulation was run for a comoving volume $V=[215.04\,{\rm Mpc}]^3$ (using a $3072^3$ grid) with a spatial resolution of $70\,{\rm kpc}$ and a mass resolution $1.09 \times 10^8 \, M_{\sun}$. We use the $N$-body particle positions to generate the \HI density field, and particle velocities to generate the velocities of \HI particles. Halos were identified using a linking length of $0.2$ times the mean inter-particle separation. We also require a halo to have a minimum of $10$ dark matter particles which corresponds to a minimum halo mass $M_{\rm halo, \, min}=1.09 \times 10^9 \, M_{\sun}$ (see e.g. \citealt{choudhury09b} for a detailed justification for this halo and particle mass threshold). The number of ionizing photons emitted by a source is assumed to be proportional to the mass of its host halo with a dimensionless proportionality constant $N_{\rm ion}$. 
The reionization process was simulated on a coarser $384^3$ grid with spacing $0.56\,{\rm Mpc}$ using the density fields and the ionizing photon fields. We determine whether a grid point is completely ionized or not by smoothing the hydrogen density fields and the ionizing photon fields using spheres of different radii starting from the grid spacing to $R_{\rm mfp}$. Here, $R_{\rm mfp}$ is a free parameter in the simulations, and analogous to the mean free path of the ionizing photons. A grid point is considered to be completely ionized if for any smoothing radius, the smoothed photon density exceeds the smoothed hydrogen density at that grid point.

For our fiducial model, we assume that the universe is 50\% ionized (the mass averaged neutral fraction $\xb \approx 0.5$) by $z=8$, $R_{\rm mfp}= 20$\,Mpc \citep{songaila10} and $N_{\rm ion}=23.2$. We also ensure that reionization ends by $z \sim 6$ and that the Thomson scattering optical depth is consistent with \citet{planck16}. 

\begin{figure*}
  \includegraphics[width=1\textwidth,angle=0]{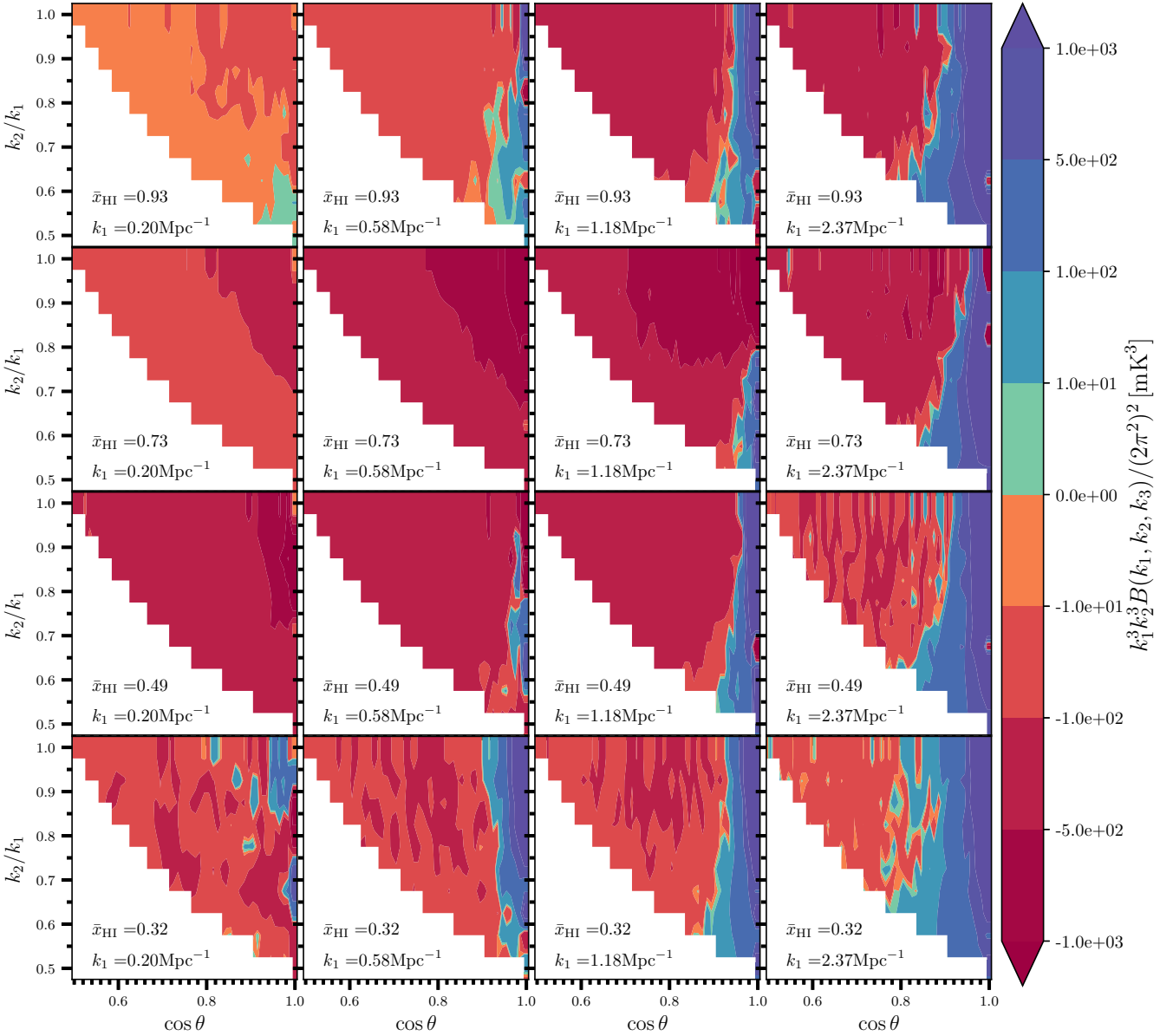}
  \caption{The real space bispectra for all unique triangle configurations at four different stages of the EoR and for four different $k_1$ modes.}
  \label{fig:bispec_real}
\end{figure*}
\begin{figure*}
  \includegraphics[width=0.95\textwidth,angle=0]{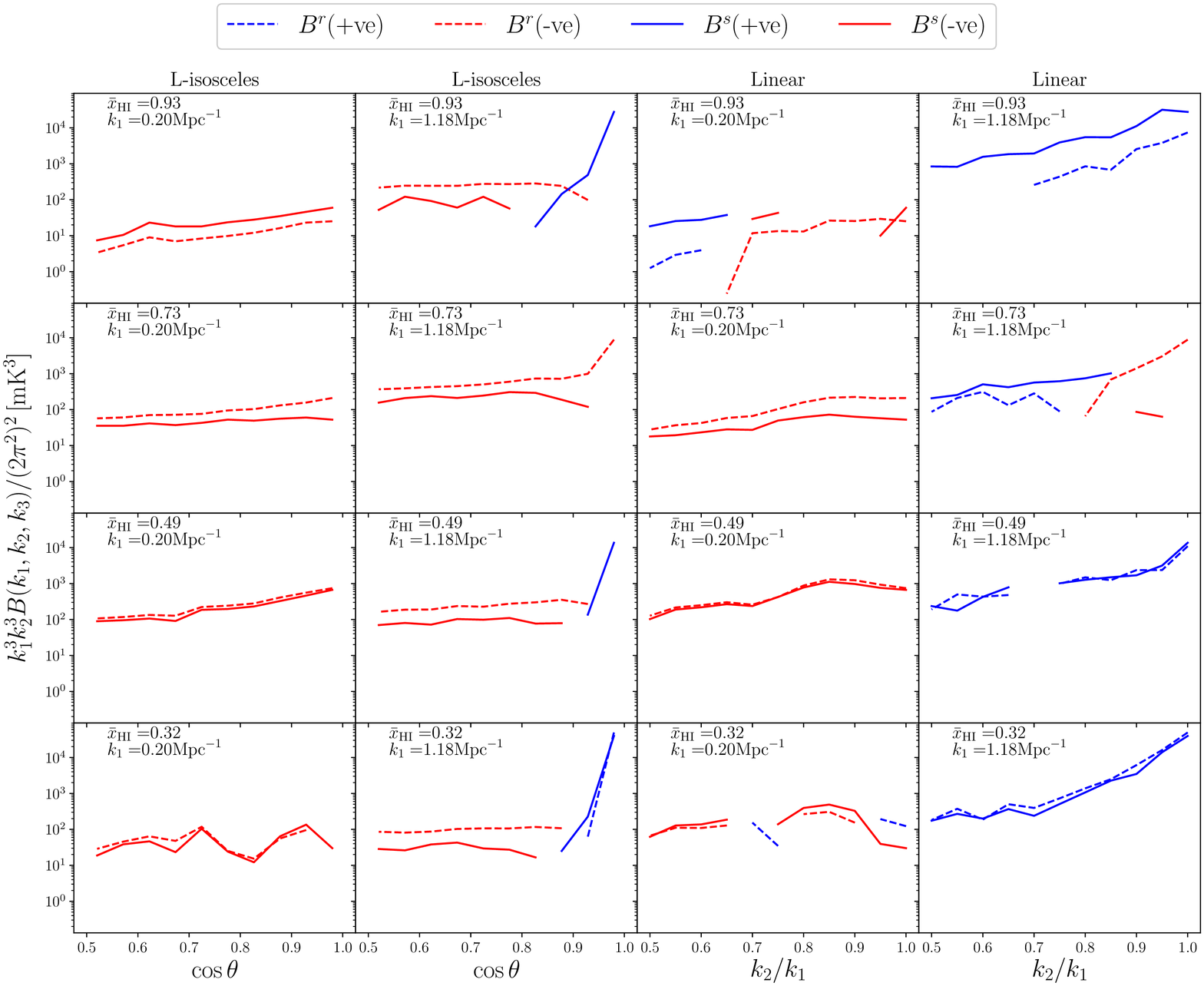}
  \caption{The bispectra in real ($B^r$) and redshift space ($B^s$) for the limiting values of $k$-triangle parameters. The solid lines and dashed lines represent bispectra in redshift and real space respectively. The red and blue colours represent negative and positive values of the bispectra respectively. The bispectra are shown at four different stages of the EoR and for two different $k_1$ modes (small and large).}
  \label{fig:bispec_line}
\end{figure*}
\section{Results}
\label{sec:results}
There are various ways in which one could normalize the bispectrum. One popular normalization method, mostly used in the analysis of the large scale structures of the Universe, is to use the reduced bispectrum $Q$ defined as $Q(k_1, k_2, k_3) =  B(k_1, k_2, k_3)/\big[  P(k_1) P(k_2)+P(k_2) P(k_3)+P(k_3) P(k_1) \big]$. This particular normalization approach is motivated by the linear perturbation theory of the density fluctuations. In the scenarios where the target field has weak non-Gaussianity, the linear perturbation theory can provide a very good approximation of the field bispectrum using different combinations of the products of the field power spectrum (\citet{Groth&Peebles1977, Fry&Peebles1978}). However, the EoR 21-cm field is expected to be highly non-Gaussian (with a significant evolution with cosmic time) in nature. Thus, this approach  (i.e. via $Q(k_1, k_2, k_3)$) of normalizing the bispectrum would not be ideal. The second approach of normalizing the bispectrum could be via $b(k_1, k_2, k_3) =  B(k_1, k_2, k_3)/\sqrt{ (k_1, k_2, k_3)^{-1} P(k_1) P(k_2) P(k_3)}$. There is an widespread use of this approach in the domain of signal processing and it was introduced by \citet{Brillinger1967}. This has also been used for the analysis of 21-cm bispectrum from the cosmic dawn by \citet{watkinson19}. As argued by \citet{watkinson19} and others this particular normalization of bispectrum (via $b(k_1, k_2, k_3)$) will effectively wash out the magnitude of the bispectrum and will retain only the phase coupling between different $k$ modes. However, in this paper we are interested in exploring all possible features (both magnitude and sign) and their evolution (across cosmic time and size and shape of $k$ triangles) of the bispectrum statistic. Additionally, if one plans to use 21-cm bispectrum as the target statistic to constrain the parameters of EoR, it will be imperative to use both the magnitude as well as the phase coupling information contained within the bispectrum to put a tighter constrain on the inferred parameter values (e.g. in case of the parameter estimation using power spectrum both the magnitude as well as the shape of the statistic is used). Therefore, throughout this paper we choose to normalize the spherically averaged bispectrum via $\big[ k_1^3 k_2^3 B(k_1, k_2,k_3)/(2\pi^2)^2\big]$, unless otherwise specified.

Following the discussion in Section \ref{sec:bispec_th} we parameterize all of the estimated bispectra from simulations with three parameters: $n$, $\cos{\theta}$ and $k_1$. As discussed in Section \ref{sec:bispec_th}, we only consider the bispectra from $k$-triangles that satisfy the uniqueness condition, $n \cos{\theta} \geq 0.5$. In discussing the results below, we focus on bispectra with $k_1 = 0.20, 0.58, 1.18, 2.37\, {\rm Mpc^{-1}}$ and designate these triangles as {\em small}, {\em intermediate}, {\em large} and {\em largest}.\footnote{Note that due to our limited simulation volume, the triangle bins with $k_1 < 0.20 \mp$ are severally affected by the sample variance. For a given $k_1$ mode, the number of closed triangles in a triangle bin for bispectrum calculation are minimum at the linear and the squeezed limit. For example for $k_1 = 0.20 \mp$ bin the minimum number of triangles at the squeezed limit is $\sim 100$ and at the linear limit it is $\sim 100-1000$, whereas for other triangle shapes it varies between $\sim 1000 - 50000$.} Furthermore, we analyze the 21-cm bispectra at four different stages of reionization corresponding to mass averaged neutral fractions $\xb = 0.93, 0.73, 0.49, 0.32$, labelled as {\em very early}, {\em early}, {\em middle} and {\em late} stages of reionization respectively.


\subsection{EoR 21-cm bispectrum in real space}
As they have not before been presented in this representation, we first show the real space EoR 21-cm bispectra for all unique triangles in the $n$-$\cos{\theta}$ space (Figure \ref{fig:bispec_real}). The first important point to note from this figure is that for almost the entire unique triangle parameter space and for all phases of reionization, the 21-cm bispectra are non-zero. This is direct evidence that the signal is highly non-Gaussian. We also notice that the magnitudes and signs of the bispectra depend on three factors, the $k$-triangle shape, the value of the $k_1$ mode and the stage of reionization. To better understand the relation of the magnitude and sign of the bispectra to these three factors, we show in Figure \ref{fig:bispec_line} the bispectra for the limiting values of $k$-triangle parameters in the $n$-$\cos{\theta}$ space (linear and L-isosceles limits for {\em small} and {\em large} $k_1$-triangles at all four stages of the EoR). 

A careful analysis of Figures \ref{fig:bispec_real} and \ref{fig:bispec_line} reveals that for {\em small} $k_1$-triangles bispectra are negative in most of the unique $n$-$\cos{\theta}$ space during the entire period of the reionization. The magnitude of the bispectra initially increases with decreasing $\xb$ and reaches its maximum value around the {\em middle} stages of the EoR, after which the magnitude decreases with decreasing $\xb$. For a fixed $\xb$, the magnitude of the bispectra increases along the L-isosceles line with increasing $\cos{\theta}$ and for linear triangles the magnitude increases with increasing $n$. The largest magnitudes are obtained for bispectra from the squeezed limit triangles. This is true for almost all stages of reionization. 

For the {\em intermediate} $k_1$-triangles the bispectra show a similar trend as for the {\em small} $k_1$-triangles. A notable exception is that the former become positive at the linear limit of triangles during the {\em late} stages of the EoR. The overall magnitude of the bispectra for almost all shapes of $k$-triangles are larger for {\em intermediate} $k_1$-triangles than for {\em small} $k_1$-triangles. 

The bispectra for the {\em large} $k_1$-triangles show even larger magnitudes. This increase is more prominent around linear triangles. In addition, these bispectra become positive in an increasingly larger area of the $n$-$\cos{\theta}$ space, around the linear triangles, for decreasing $\xb$.   

This trend of increase in the area of the $n$-$\cos{\theta}$ space  where bispectra is positive with the increase in $k_1$ magnitude is seen to continue for the {\em largest} $k_1$-triangles as well. The area of the $n$-$\cos{\theta}$ space around the linear triangles where the bispectra become positive, is even larger for this case. However, the magnitude of the bispectra for the {\em largest} $k_1$-triangles monotonically decreases with the progress of the reionization. This is a feature that is opposite of the trend seen for the {\em small}, {\em intermediate} and {\em large} $k_1$-triangles. 

The $k_1$ value, triangle shape and $\xb$ dependence of the bispectra magnitude observed in Figure \ref{fig:bispec_real} and \ref{fig:bispec_line} and discussed above can be interpreted in the following manner: the signal's non-Gaussianity is relatively small at large scales during the {\em very early} stages of the EoR. This can be understood from the fact that although the fluctuations in the 21-cm signal during these early stages are dominated by the fluctuations introduced by the $\HII$ regions, the $\HII$ regions are still very small in size. Thus the overall amplitude of fluctuations in the signal is relatively low in magnitude compared to the signal from the later stages of the EoR. When we probe bispectra for triangles with {\em intermediate, large} and {\em largest} $k_1$ modes, the magnitude of the bispectra increases as they become increasingly sensitive to the small $\HII$ regions (left to right, top two rows of the Figure \ref{fig:bispec_real}). Furthermore, the gradual increase in bispectra magnitude with decreasing $\xb$ and peaking around $\xb \sim 0.5$ is directly related to the gradual increase in the signal fluctuations (and non-Gaussianity) due to the progress of reionization until we reach a significant amount of percolation among the $\HII$ regions. Furthermore, progress in reionization changes the topology of the 21-cm signal and for $\xb \lesssim 0.5$ it is dictated by the size and distribution of the neutral regions rather than of the ionized regions. The bispectra for triangles with the {\em largest} $k_1$ modes see a decrease in magnitude with the decreasing $\xb$. This is caused by the fact that as reionization progresses the sizes and number density of the $\HII$ regions gradually increase which leads to a decrease of the signal fluctuations at smaller length scales \citep{lidz07}.

The sign reversal of the bispectra discussed above is an important phenomenon, which has been reported earlier in the context of the 21-cm signal from the EoR in \citet{majumdar18} and \citet{hutter19} and in the context of the 21-cm signal from the CD in \citet{watkinson19}. However, all of these previous studies have reported this sign change for a few specific types of triangles. Here we demonstrate the evolution in bispectra sign across all unique triangle types, $k_1$ modes and stages of reionization. 

\subsection{EoR 21-cm bispectrum in redshift space}

\begin{figure*}
  \includegraphics[width=1\textwidth,angle=0]{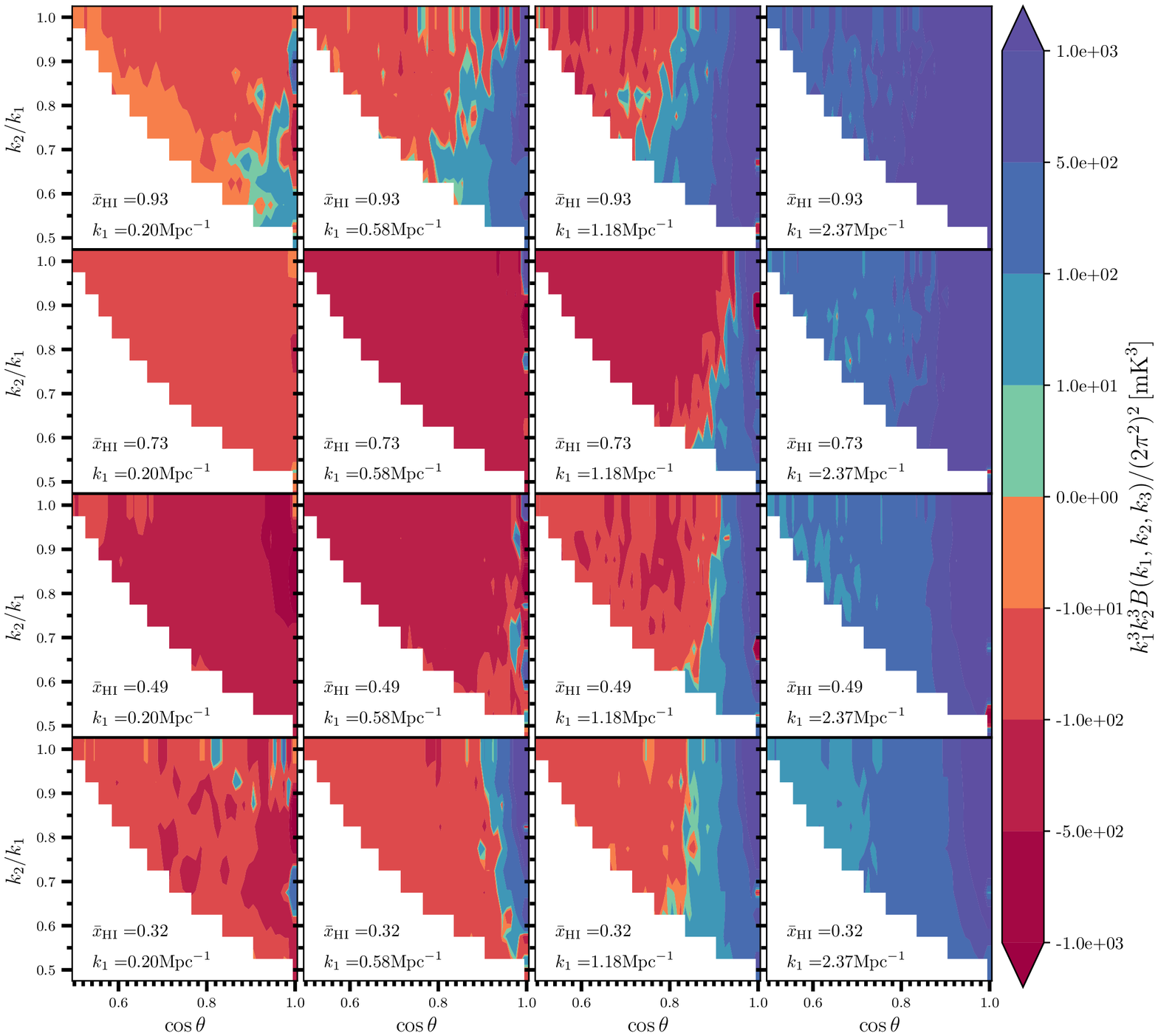}
  \caption{The redshift space bispectra for all unique triangle configurations at four different stages of the EoR and for four different $k_1$ modes.}
  \label{fig:bispec_rsd}
\end{figure*}
\begin{figure*}
  \includegraphics[width=1\textwidth,angle=0]{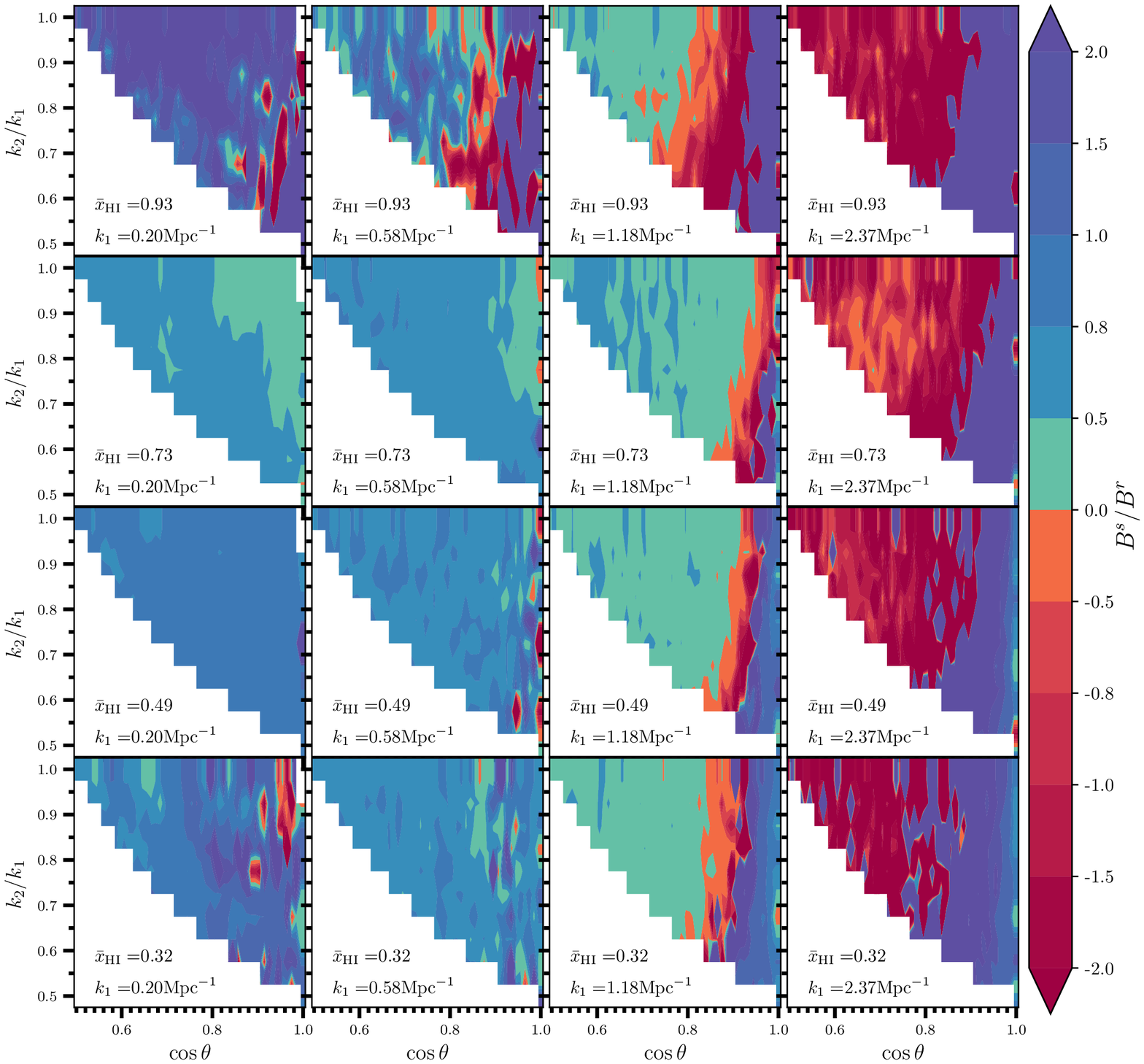}
  \caption{The ratio between the redshift space and real space bispectra for all unique triangle configurations at four different stages of the EoR and for four different $k_1$ modes.}
  \label{fig:bispec_rsd_real_ratio}
\end{figure*}
\begin{figure*}
  \includegraphics[width=1\textwidth,angle=0]{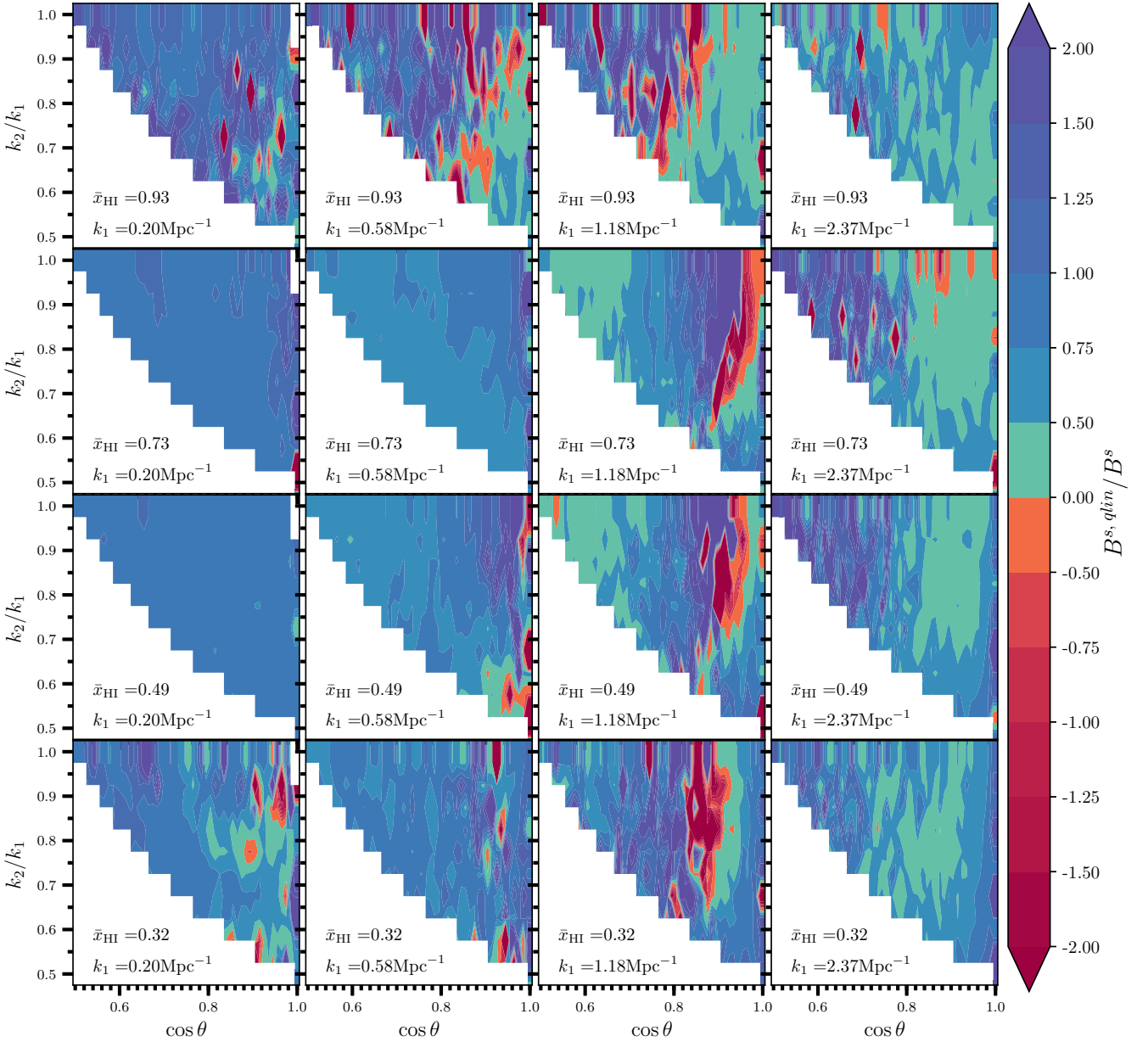}
  \caption{This figure shows $B^{\rm s,\:qlin}/B^{\rm s}$ for all unique triangle configurations at four different stages of the EoR and for four different $k_1$ modes.}
  \label{fig:bispec_rsd_reconst_orig_rsd_ratio}
\end{figure*}
\begin{figure*}
  \includegraphics[width=0.95\textwidth,angle=0]{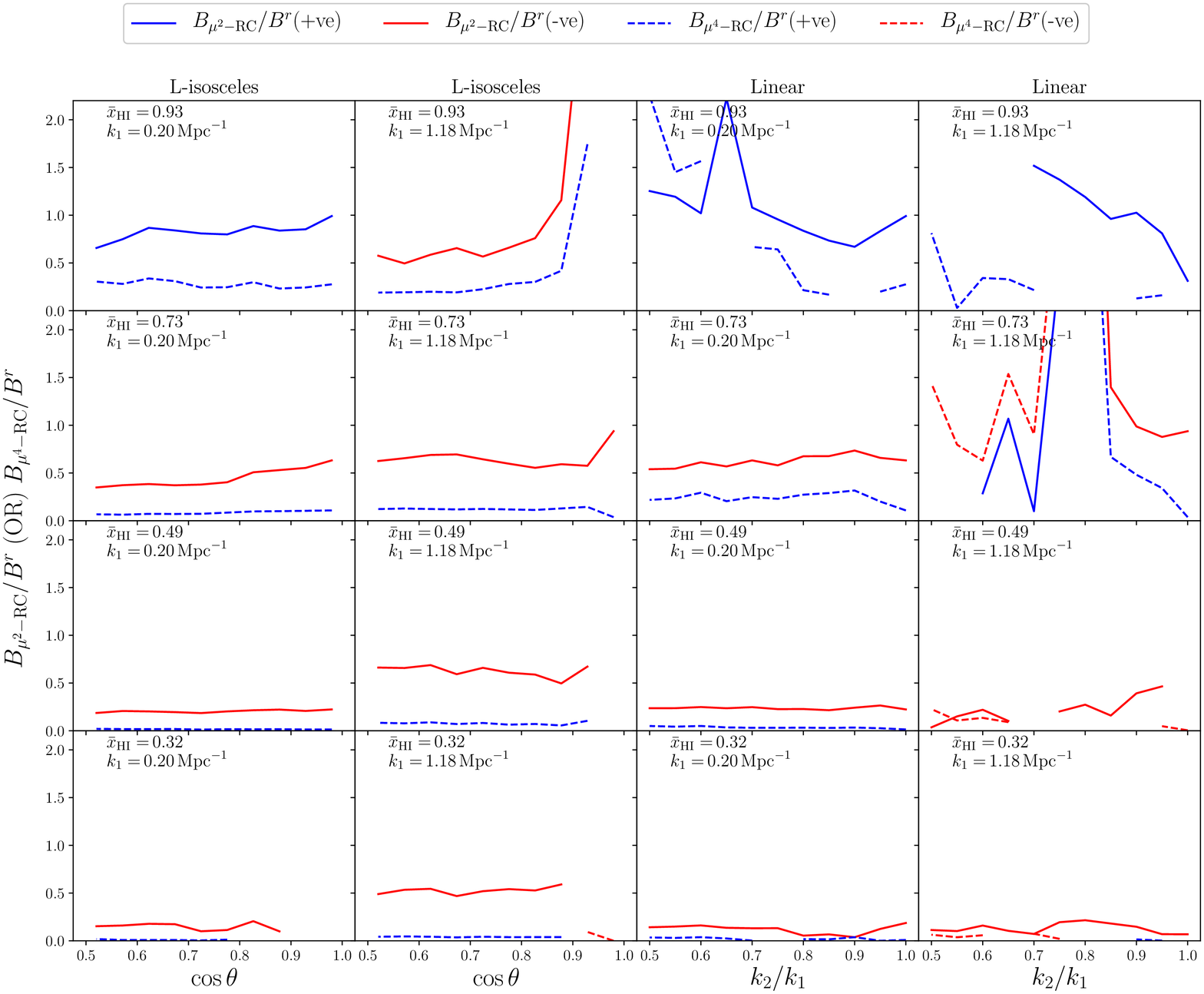}
  \caption{The ratios $B_{\mu^2-{\rm RC}}/B^r$ and $B_{\mu^4-{\rm RC}}/B^r$ for the limiting values of $k$-triangle parameters in the $n$-$\cos{\theta}$ space. The solid lines and dashed lines show the ratios $B_{\mu^2-{\rm RC}}/B^r$ and $B_{\mu^4-{\rm RC}}/B^r$ respectively. The red and blue colours represent negative and positive values of the ratios respectively. The ratios are shown at four different stages of the EoR and for two different $k_1$ modes ({\em small} and {\em large}).}
  \label{fig:bispec_rsd_RC}
\end{figure*}
\begin{figure*}
  \includegraphics[width=1\textwidth,angle=0]{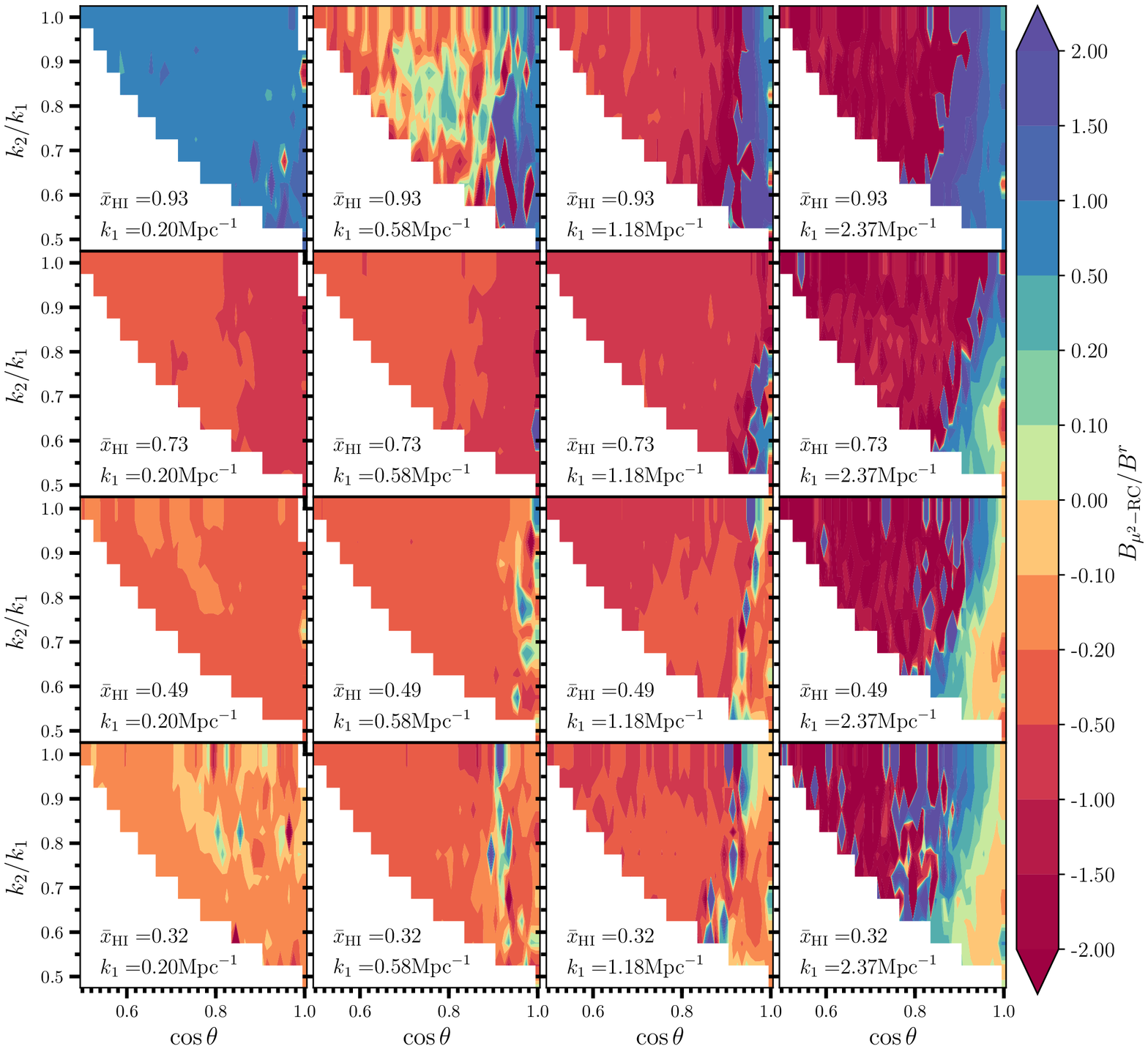}
  \caption{The ratio between the ${B_{{\mu^2}-{\rm RC}}}$  and $B^r$  for all unique triangle configurations at four different stages of the EoR and for four different $k_1$ modes.}
  \label{fig:B_mu2_to_Br}
\end{figure*}
\subsubsection{Impact of the RSD on the 21-cm bispectra}  
Next, we quantify the impact of the RSD on the monopole moment of the 21-cm bispectra. Figure \ref{fig:bispec_rsd} shows the spherically averaged redshift space bispectra for all unique $k$-triangles in the $n$-$\cos{\theta}$ space. Similar to the case of the real space bispectra, in Figure \ref{fig:bispec_line} we also show the redshift space bispectra for the limiting values of $k$-triangle parameters in the $n$-$\cos{\theta}$ space (linear and L-isosceles limits for {\em small} and {\em large} $k_1$-triangles at all four stages of the EoR).

A quick qualitative visual comparison of  Figures \ref{fig:bispec_real}, \ref{fig:bispec_rsd} and  \ref{fig:bispec_line} reveals many of the important effects RSD has on the signal bispectra. The first is that both the magnitude and the sign of the bispectra are affected. For {\em small} $k_1$-triangles and at {\em very early} stages of the EoR, RSD manifests itself through a large boost in the magnitude values. Similarly, the magnitude of the bispectra for {\em large} $k_1$-triangles also get boosted at this stage and in addition they also show a sign change. As reionization progresses, we see an opposite effect on the magnitudes, as they are smaller for the RSD case for both {\em small} and {\em large} $k_1$-triangles. It is further clearly visible from Figure \ref{fig:bispec_rsd} that the area in the $n$-$\cos{\theta}$ space where the bispectra become positive for {\em large} $k_1$-triangles, is larger when RSD is applied. For the {\em largest} $k_1$-triangles, RSD makes the bispectra positive for the entire unique $n$-$\cos{\theta}$ space at all stages of the EoR.
 
To quantify the impact of the RSD in detail, we show the ratio of spherically averaged bispectra in redshift and real space, i.e. $B^s/B^r$ in Figure \ref{fig:bispec_rsd_real_ratio}. We discuss this ratio in order of increasing $k_1$ values. Starting with the {\em small} $k_1$-triangles we notice that during the {\em very early} stages of the EoR, RSD enhances the magnitude of the bispectra by more than $50\%$ in almost the entire unique $k$-triangle space. However, as reionization progresses ({\em early} stages), RSD reduces the amplitudes of the bispectra, by $\sim 50\%$ around the squeezed and linear limit and by $\sim 20-50\%$ for the other $k_1$-triangles. During the {\em middle} and {\em late} stages of the reionization, RSD continues to reduce the amplitude of the bispectra but to a smaller degree ($\leq 20\%$). 
 
For {\em intermediate} $k_1$-triangles the bispectra witness a stronger impact (both in magnitude and sign) of the RSD. At the {\em very early} stages of the EoR, the $B^s$ has a different sign than $B^r$ for the $\cos{\theta}$ range $0.8 \lesssim \cos{\theta} \lesssim 0.9$. At these scales the magnitude of $B^s$ is enhanced by $\sim 50-100\%$ for the linear $k$-triangles for $0.9 \lesssim \cos{\theta} \lesssim 1.0$ for the entire range of $n$. As reionization progresses ({\em early} stages) the magnitude of $B^s$ reduces with respect to $B^r$ by more than $\sim 50\%$ in the vicinity of linear $k$-triangles with $0.6 \lesssim n \lesssim 1.0$ and $0.9 \lesssim \cos{\theta} \lesssim 1.0$.  During the {\em middle} and {\em late} stages of the EoR the reduction in bisepctra amplitude due to RSD remains within $\leq 20\%$ in almost the entire unique $n$-$\cos{\theta}$ space.
 
The {\em large} $k_1$-triangle bispectra are even more sensitive to the effect of redshift space distortions. This is maybe due to the fact that non-linear features of the signal are more prominent at small length scales. The sign of the $B^s$ is the opposite of that of the $B^r$ for triangles with $0.7 \lesssim \cos{\theta} \lesssim 0.9$ during the {\em very early} stages of the EoR. This range of $\cos{\theta}$, for which a sign difference is observed shifts to higher values of $\cos{\theta}$ (i.e.\ $0.9 \lesssim \cos{\theta} \lesssim 1.0$) as the reionization transitions from the {\em very early} to the {\em early} stages. As reionization progresses further (i.e. {\em middle} and {\em late} stages) this $\cos{\theta}$ range (where the sign difference is observed) again shifts towards smaller values (i.e.\ $0.8 \lesssim \cos{\theta} \lesssim 0.9$). The magnitude of the bispectra are also affected by the RSD for {\em large} $k_1$-triangles. At the {\em very early} stages of the EoR and for $0.5 \lesssim \cos{\theta} \lesssim 0.8$ the RSD decreases the magnitude by more than $\sim 50\%$ and for $0.8 \lesssim \cos{\theta} \lesssim 1.0$ the RSD increases the magnitude by more than $\sim 50\%$. During the later stages ({\em early, middle} and {\em late} stages of the EoR) the magnitude of $B^s$ decreases by more than $\sim 50\%$, except for the region in the $n$-$\cos{\theta}$ space where the ratio $B^s/B^r$ changes sign.
 
The RSD has its maximum impact both in terms of sign and magnitude on the bispectra for the {\em largest} $k_1$-triangles. As already noted above, for these triangles $B^s$ is positive for the entire unique $n$-$\cos{\theta}$ space during the entire period of reionization. Therefore the sign of the $B^s$ is the opposite of that of $B^r$ for $0.5 \lesssim \cos{\theta} \lesssim 0.85$ (Figure \ref{fig:bispec_rsd_real_ratio}). The magnitude also changes by roughly $\sim 50\%$ in this region of parameter space. Furthermore, for $0.85 \lesssim \cos{\theta} \lesssim 1.0$ the magnitude of $B^s$ enhances by more than $\sim 50\%$.

Overall it can be concluded that, the impact of RSD on the magnitude of the bispectrum (for any type of $k$-triangle and for any $k_1$-mode) is minimum when reionization is roughly half way through.

\subsubsection{Interpretation of the redshift space bispectra using the quasi-linear model}
\label{sec:interpretation}
Several toy models for the size and distribution of the ionized regions have been used to interpret and explain the real space EoR 21-cm bispectrum \citep{bharadwaj05a, majumdar18}. However, building such a toy model for the redshift space 21-cm  bispectrum is difficult as the redshift space distortions changes the signal in a very complex manner. A matter density field, when subject to the redshift space distortions, changes in the following way: a spherical overdensity region will attract more matter radially from its surroundings, which will make it appear squashed along the LoS of the observer. Whereas a spherical low density region will radially loose matter to its neighbours and thus will appear elongated along the LoS of the observer. If we consider an inside-out model for reionization (which is the case for our simulated 21-cm signal), there the highest matter overdensity regions will ionize first and after which the lower matter density regions will follow suit. Due to the redshift space distortions these early ionized regions (which are located on highest matter overdensity regions) will appear squashed along the LoS and the early neutral regions will appear stretched along the LoS in the 21-cm maps.  However, this simplistic picture is somewhat valid at the very early stages of the EoR and for spherical ionized regions. As reionization progresses, more and more lower matter density regions get ionized and the ionized regions themselves also start merging together then this simplistic picture can no longer describe the 21-cm fluctuations and the impact of the redshift space distortions on it.

To interpret the features observed in the redshift space bispectra we therefore turn to the quasi-linear model for $B^s$ expressed by Equation \eqref{eq:B_qlin1}. However, before interpreting the results using this model we first check in which regions of the $n$-$\cos{\theta}$ space, for what $k_1$-values and for what stages of the EoR it provides a good approximation to $B^s$. Figure \ref{fig:bispec_rsd_reconst_orig_rsd_ratio} shows the ratio of the bispectra estimated from Equation \eqref{eq:B_qlin1} (using the simulated $\rho_{\rm H}$ and $\rho_{\HI}$ fields) and the bispectra estimated from the simulated redshift space 21-cm brightness temperature fields, i.e.\ $B^{\rm s,\, qlin}/B^{\rm s}$. We observe that the quasi-linear model provides a very good estimate (with deviations of less than $10$ percent from the actual $B^s$) in almost the entire $n$-$\cos{\theta}$ space and for the entire period of reionization for {\em small} and {\em intermediate} $k_1$-triangles. The quasi-linear predictions for bispectra for {\em large} and {\em largest} $k_1$-triangles show somewhat higher deviations ($\sim 10$--50 percent). However, even for this group of triangles in most of the scenarios, the quasi-linear model is able to predict the magnitude and the sign of the $B^s$ reasonably well.

As discussed in Section \ref{subsec:quasi}, the R.H.S. of Equation \eqref{eq:B_qlin1} introduces seven correction (RC) terms to the real space bispectra as a model for the redshift space bispectra. We divide these seven RC terms into three groups - $B_{\mu^2-{\rm RC}}$, $B_{\mu^4-{\rm RC}}$ and $B_{\mu^6-{\rm RC}}$ (see Section \ref{subsec:quasi} for details). Next, we try to identify which one among these three groups of RC terms is the dominant one. To this end we plot in Figure \ref{fig:bispec_rsd_RC} the ratios $B_{\mu^2-{\rm RC}}/B^r$ and $B_{\mu^4-{\rm RC}}/B^r$ for the limiting values of $k$-triangle parameters in the $n$-$\cos{\theta}$ space (linear and L-isosceles) for {\em small} and {\em large} $k_1$-triangles. The advantage of plotting this ratio is that apart from quantifying the relative magnitudes of the RC terms, the sign of the ratio will tell us whether a given RC term is contributing with the same sign (positive ratio) or with the opposite sign (negative ratio) of the real space bispectra. We do not plot the ratio 
$B_{\mu^6-{\rm RC}}/B^r$ as its magnitude is negligible compared to the ratios $B_{\mu^2-{\rm RC}}/B^r$ and $B_{\mu^4-{\rm RC}}/B^r$ at almost all stages of the EoR and for almost all of the studied triangle types.

The figure clearly shows that for these limiting triangle shapes the dominant RC term is $B_{\mu^2-{\rm RC}}$ among $B_{\mu^2-{\rm RC}}$ and $B_{\mu^4-{\rm RC}}$. At the {\em very early} stages of the EoR and for {\em small} $k_1$-triangles $B_{\mu^2-{\rm RC}}/B^r$ is positive and its magnitude varies in the range 0.6 -- 2.0, whereas in the same regime $B_{\mu^4-{\rm RC}}/B^r$ is also positive and its magnitude varies in the range $0.1-0.6$ for most the triangle shapes. This explains why we see a boost in the bispectra due to the RSD during the {\em very early} stages of the EoR and for {\em small} $k_1$-triangles (see Figure \ref{fig:bispec_line}). During all the later stages of the EoR (i.e. {\em early, middle} and {\em late} stages) for the {\em small} $k_1$-triangles $B_{\mu^2-{\rm RC}}/B^r$ is negative and its magnitude reduces as reionization progresses. However, $B_{\mu^4-{\rm RC}}/B^r$ is positive for the same triangles but has a much smaller magnitude compared to $B_{\mu^2-{\rm RC}}/B^r$ at the {\em early} stages of the EoR. During the {\em middle} and {\em late} stages of the EoR $B_{\mu^4-{\rm RC}}/B^r$ for same triangles is negligible. This explains why we see a reduction in the magnitude of the redshift space bispectra (with respect to the real space bispectra) for {\em small} $k_1$-triangles with the progress of reionization. This is also the reason why during the {\em late} stages of the EoR eventually $B^s \sim B^r$. A somewhat similar behaviour of $B_{\mu^2-{\rm RC}}/B^r$ and $B_{\mu^4-{\rm RC}}/B^r$ is observed for {\em large} $k_1$-triangles as well. 

Next, to understand how the $B_{\mu^2-{\rm RC}}$ term shapes the redshift space bispectra for all types of triangles and for the entire period of the reionization, we show in Figure \ref{fig:B_mu2_to_Br} the ratio $B_{\mu^2-{\rm RC}}/B^r$ for the entire unique $n$-$\cos{\theta}$ space following the convention of Figure \ref{fig:bispec_real}. In line with what we have observed for the limiting shapes of the triangles in Figure \ref{fig:bispec_rsd_RC}, Figure \ref{fig:B_mu2_to_Br} shows that for the {\em small} $k_1$-triangles during the {\em very early} stages of the EoR $B_{\mu^2-{\rm RC}}/B^r \sim 1.0$ and also has a positive sign. The $B_{\mu^4-{\rm RC}}/B^r$ and $B_{\mu^6-{\rm RC}}/B^r$ ratio for the same triangles (shown in Figures \ref{fig:B4_to_Br} and \ref{fig:B6_to_Br}) have a magnitude in the range $0.0 - 0.1$ with opposite signs with respect to each other. This explains why we observe a boost in the magnitude of redshift space bispectra in the entire unique $n$-$\cos{\theta}$ space at this stage. For the later stages of the EoR the $B_{\mu^2-{\rm RC}}/B^r$ becomes negative and also shows a decrease in magnitude with decreasing $\xb$. However, it still remains the dominant RC term and can explain the decrease in amplitude of the redshift space bispectra (compared to real space bispectra) during the {\em early} and {\em middle} stages of the EoR until $B^s \sim B^r$ at the {\em late} stages of the EoR. 

For the {\em intermediate} $k_1$-triangles at the {\em very early} stages of the EoR the magnitude of $B_{\mu^2-{\rm RC}}/B^r$ falls roughly in the range 0.2--1.0 and has a negative sign (Figure \ref{fig:B_mu2_to_Br}). $B_{\mu^4-{\rm RC}}/B^r$ ranges from 0.5 to 1.0 with mostly a positive sign (Figure \ref{fig:B4_to_Br}) and $B_{\mu^6-{\rm RC}}/B^r$ varies from 0.1 to 1.0 with mostly a negative sign (Figure \ref{fig:B6_to_Br}). Hence in this regime the contribution of $B_{\mu^2-{\rm RC}}$ is comparable with the other two RC terms. Further, as the two RC terms with higher powers of $\mu$ have  competing signs, the RSD bispectra show a fluctuating sign when compared with the real space bispectra (see Figure \ref{fig:bispec_rsd_real_ratio}) for this regime. As reionization progresses (i.e.\ {\em early, middle} and {\em late} stages) all three correction terms - $B_{\mu^2-{\rm RC}}/B^r$, $B_{\mu^4-{\rm RC}}/B^r$ and $B_{\mu^6-{\rm RC}}/B^r$ follow a somewhat similar behaviour as for the {\em small} $k_1$-triangles case described above and thus the redshift space bispectra also follow the suit.

At the {\em very early} stages of the EoR the $B_{\mu^2-{\rm RC}}/B^r$ for {\em large} $k_1$-triangles shows a variation in magnitude with triangle shape. The magnitude varies in the range 1.0--2.0 and it gradually increases as we approach the linear regime of triangles along the $\cos{\theta}$ axis. It also changes sign from negative to positive around $\cos{\theta} \sim 0.9$. The $B_{\mu^4-{\rm RC}}/B^r$ ratio  shows a similar variation in magnitude ranging from 0.2 to 1.0 and the sign changes from positive to negative around $\cos{\theta} \sim 0.9$. Similarly, the magnitude of $B_{\mu^6-{\rm RC}}/B^r$ ranges from 0.1 to 1.5 and its sign changes from negative to positive around $\cos{\theta} \sim 0.9$. This boosts the magnitude of the redshift bispectra significantly around the linear regime of triangles and also ensures it is positive in the same region of the $n$-$\cos{\theta}$ space. The overall magnitude of all three RC terms decreases as reionization progresses. However, the trend in the variation of magnitude with the triangle shape remains more or less the same. This results in a overall decrease in the magnitude of the redshift space bispectra with the values around the linear regime of triangles and a sign change in the redshift space bispectra around $\cos{\theta} \sim 0.9$.

For the {\em largest} $k_1$-triangles the magnitude of   $B_{\mu^2-{\rm RC}}/B^r$ varies in the range 0.5--2.0 and changes its sign from negative to positive around $\cos{\theta} \sim 0.8$ at almost all stages of the EoR. Similarly, the magnitude of $B_{\mu^4-{\rm RC}}/B^r$ range from 0.1 to 0.5 with a sign change from positive to negative around the same $\cos{\theta}$ value as above. The relative magnitude of $B_{\mu^6-{\rm RC}}/B^r$ with respect to the other two RC terms is somewhat negligible in this regime ($\leq 0.1$). This effectively ensures that the redshift space bispectra are positive at all stages of the EoR. The magnitude of the bispectra increases with increasing values of $\cos{\theta}$ and reaches a maximum around the linear limit of triangles (Figure \ref{fig:bispec_rsd}).      
\section{Summary and Conclusions}
\label{sec:summary}

In this article we have presented a comprehensive study of the spherically averaged EoR 21-cm signal bispectra, which are a probe of the non-Gaussianity present in the signal. This work is the first of its kind, as it quantifies the EoR 21-cm bispectra for all possible unique $k$-triangles in the triangle parameter ($n$-$\cos{\theta}$) space using an ensemble of simulated signals. All previous efforts in estimating the EoR 21-cm bispectra were less complete as they were limited to a few specific kind of $k$-triangles. This article is also the first to quantify the impact of redshift space distortions on the signal bispectra. 

We find that the 21-cm bispectra are non-zero for most of the triangle parameter space and during almost the entire period of the reionization. This strongly establishes that the EoR 21-cm signal is highly non-Gaussian. Our findings can be further summarized as below:

\begin{itemize}

    \item The magnitude of both the real and redshift space signal bispectra (in the entire triangle parameter space) initially increases with decreasing $\xb$ for all $k_1$-triangles having values $k_1 \lesssim 1.0 \mp$. They achieve their maximum approximately for $\xb \sim 0.5$ after which they decrease with decreasing $\xb$. This is due to the fact that the signal fluctuations increase gradually with the increasing sizes of the $\HII$ regions and peak at the mid-point of reionization.
    
    \item The 21-cm bispectra in both real and redshift space show a gradual increment in the magnitude as we go from smaller to larger $k_1$ values (the largest arm in the $k$-triangle) with $k_1 \lesssim 1.0 \mp$ and for a fixed $\xb$. This seems to be a signature of the impact of the \HII region size distribution at a given stage of the EoR.
    
    \item The sign of the EoR 21-cm bispectra, an important feature of this statistic, is negative for most of the $n$-$\cos{\theta}$ space for the real space signal (across all $k_1$-triangles and for almost all $\xb$ values). It is positive only in the limit and vicinity of squeezed ($n \sim 1.0$ and $\cos{\theta} \sim 1.0$) and linear (i.e. $n \sim 0.5-1.0$ and $\cos{\theta} \sim 1.0$) $k_1$-triangles. The region of positive bispectra in the $n$-$\cos{\theta}$ space increases in area as we move from smaller to larger $k_1$-triangles. Another important point to note is that the magnitude of the bispectra reaches its maximum for the squeezed and linear triangles.
    
    \item The redshift space distortions affect the bispectra for all unique $k$-triangles significantly, both in terms of magnitude and sign, during the entire period of the reionization. The impact due to RSD on the bispectra magnitude is larger (as large as $\sim 100 \%$) during the early stages of the EoR ($\xb \gtrsim 0.7$) for triangles with small and intermediate $k_1$ modes ($k_1 \lesssim 0.6 \mp$). The RSD have a smaller impact (at most $\sim 50\%$) on the bispectra magnitude during the later stages of the EoR ($\xb \lesssim 0.7$) for the same $k_1$-triangles. 
    
    \item The gradual change in the sign of the bispectra is most prominent when one analyses the bispectra across small to large $k_1$-triangles for any given stage of reionization. The signal bispectra in redshift space are mostly negative for all unique triangles with $k_1 \sim 0.2 \mp$ (small) all stages of the EoR. As we move from smaller to larger $k_1$ triangles at any give stage of reionization, an area with positive bispectra starts to appear close to the $\cos{\theta} \sim 1$ line which increases in size the larger $k_1$ becomes. In case of the largest $k_1$-triangles discussed here ($k_1 \sim 2.4 \mp$), the bispectra have positive sign for the entire $n$-$\cos{\theta}$ parameter space. A similar trend in the evolution of the sign is observed for the real space signal as well. However, the area in the $n$-$\cos{\theta}$ space where bispectra are positive is much smaller in real space than in redshift space. This is true for all stages of reionization and for all $k_1$-triangles.   
 
    \item The RSD have their maximum impact on the larger $k_1$-triangle bispectra ($k_1 \gtrsim 0.6 \mp$). It enhances the magnitude of the signal bispectra for $k_1 \sim 1.0 \mp$ triangles by $\sim 100\%$, in the region of the $n$-$\cos{\theta}$ space where $B^s$ and $B^r$ have opposite signs. The bispectra for other unique triangles in the $n$-$\cos{\theta}$ (for the same $k_1$ values) space experience a decrease in magnitude, sometimes as low as $\sim 80$ percent, due to the RSD. For $k_1 \sim 2.4 \mp$ triangles, the RSD change the magnitudes by at least $\sim 100\%$ in most of the unique $n$-$\cos{\theta}$ space. Additionally, it also changes the signs within $\cos{\theta}$ range $0.5 \lesssim \cos{\theta} \lesssim 0.85$. 
   
    \item The quasi-linear model (Equation \eqref{eq:B_qlin1}) provides a very good prediction and physical interpretation for the redshift space EoR 21-cm bispectra (with $\leq 10\%$ uncertainties) in the entire unique $n$-$\cos{\theta}$ space for $k_1 \leq 0.6 \mp$ triangles during the early stages of the EoR ($\xb \geq 0.5$). These predictions deviate more ($\geq 20$ percent) from the simulated $B^s$ as we move towards triangles with larger $k_1$ modes ($k_1 \geq 0.6 \mp$) and to the later stages of the EoR. We have further established that among the three groups of the RC terms shown in Equation \eqref{eq:B_qlin1}, mainly the group $B_{\mu^2-{\rm RC}}$ dominates in shaping the redshift space 21-cm bispectra. This group  contains three cross-bispectra which are   $[\overline{\mu_1^2}]_0^0B_{\Delta_{\rho_{\rm H}}, \Delta_{\rho_{\HI}}, \Delta_{\rho_{\HI}}}$, $[\overline{\mu_2^2}]_0^0B_{\Delta_{\rho_{\HI}}, \Delta_{\rho_{\rm H}}, \Delta_{\rho_{\HI}}}$ and $[\overline{\mu_3^2}]_0^0B_{\Delta_{\rho_{\HI}}, \Delta_{\rho_{\HI}}, \Delta_{\rho_{\rm H}}}$. The other RC terms do not have a similar impact in shaping the redshift space signal bispectra.  
 \end{itemize}
 
The analysis of the simulated EoR 21-cm signal bispectra presented here is quite comprehensive in nature. It establishes that the impact of redshift space distortions on the signal bispectrum is significant, both in terms of bispectra magnitude and sign. Thus it is important to take into account the effect of the RSD for any interpretation of the signal bispectra.

However, in this article, we only focus on the impact of RSD on the spherically averaged bispectra of the signal. Ideally, one should decompose the direction dependent signal bispectra into an orthonormal basis vector space to accurately quantify any line-of-sight anisotropy present in the signal (e.g. \citealt{bharadwaj20}). 
 
We would like to point out that our analysis does not consider several other effects which are unavoidable in any radio interferometric observations of the signal. Among them, one that is very important, is the light cone effect, which arises due to the time evolution of the signal along the line-of-sight. This constitutes another source of line-of-sight anisotropy in the signal apart from the redshift space distortions, although it generally of a lower magnitude \citep{datta12, datta14}.

Furthermore, the analysis presented here does not take into account the corruption of the observed data due to the presence of residual foregrounds. A proper foreground removal or avoidance is essential to extract the cosmic signal using the bispectrum (e.g. \citealt{watkinson20}). Additionally, we have not presented the detectability of the signal bispectra for any upcoming or presently operational radio interferometers by considering the level of noise and other systematics.

We have considered only one model of reionization for our analysis of the 21-cm bispectrum here. However, as discussed earlier, the nature of the bispectra depends on the topology of the 21-cm field, which in turn depends on the properties of the ionizing sources as well as the properties of the IGM (see e.g. \citealt{majumdar16}).

Lastly, this analysis also does not consider the period of cosmic history when the effect of spin temperature fluctuations on the signal is significant. However, through the use of the $(k_1, n, \cos{\theta})$ parameter space we have shown how all possible triangle configurations can be studied, thus establishing a framework which we intend to use in future studies addressing these various issues.

\section*{Acknowledgements}
JRP and SM acknowledge financial
support from the European Research Council under ERC grant number 638743-FIRSTDAWN. SM and GM acknowledge financial support from the ASEM-DUO India 2020 fellowship. RM is grateful for financial support from the Wenner-Gren Foundations.

\section*{Data Availability}
The simulated data underlying this article will be shared on reasonable request to the corresponding author.
\bibliographystyle{mn2e} 
\bibliography{refs}
\appendix
\section{Quasi-linear components of the redshift space 21-cm bispectra}
\label{sec:app1}
\begin{figure*}
  \includegraphics[width=0.24\textwidth,angle=0]{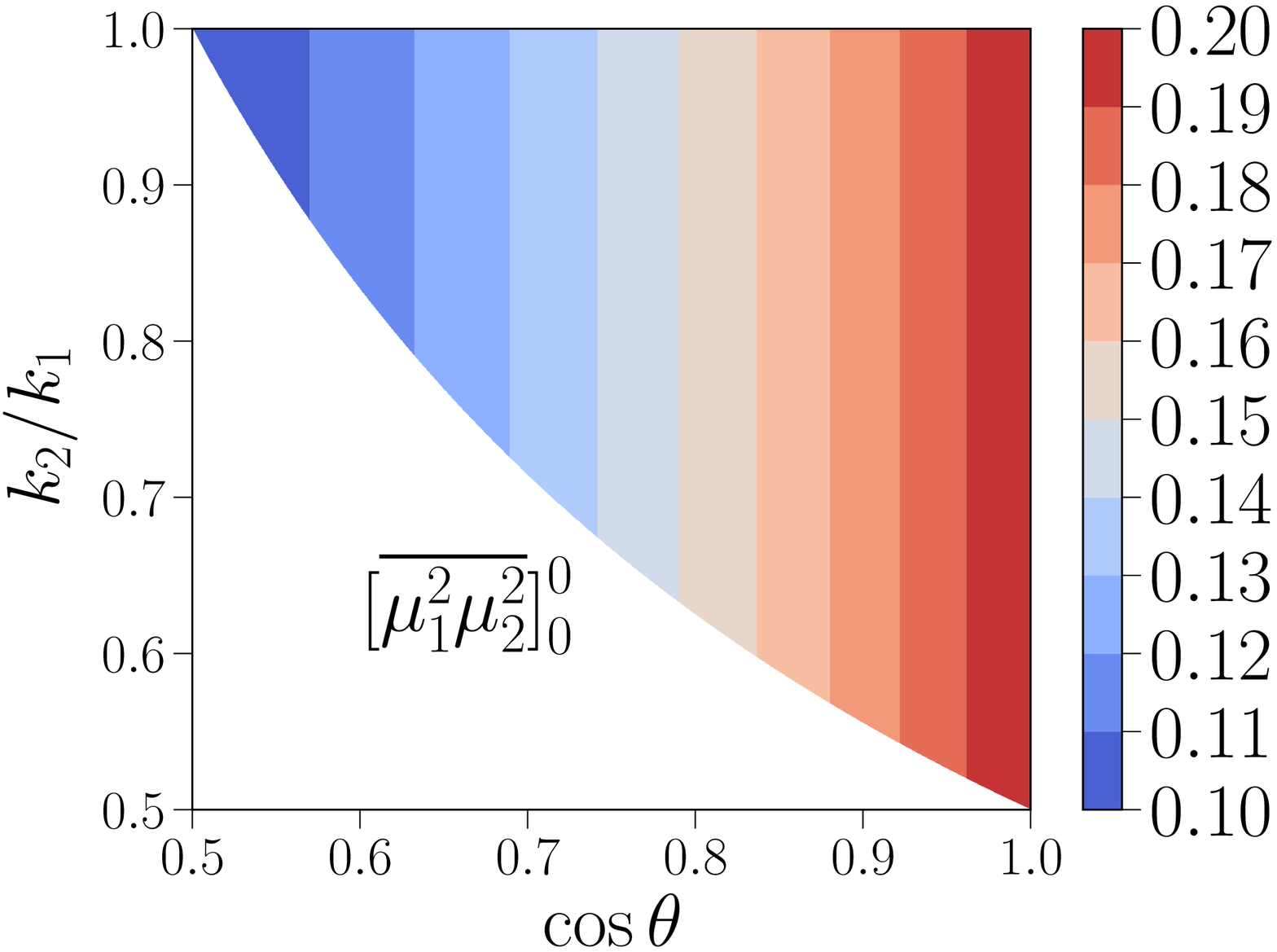}
  \includegraphics[width=0.24\textwidth,angle=0]{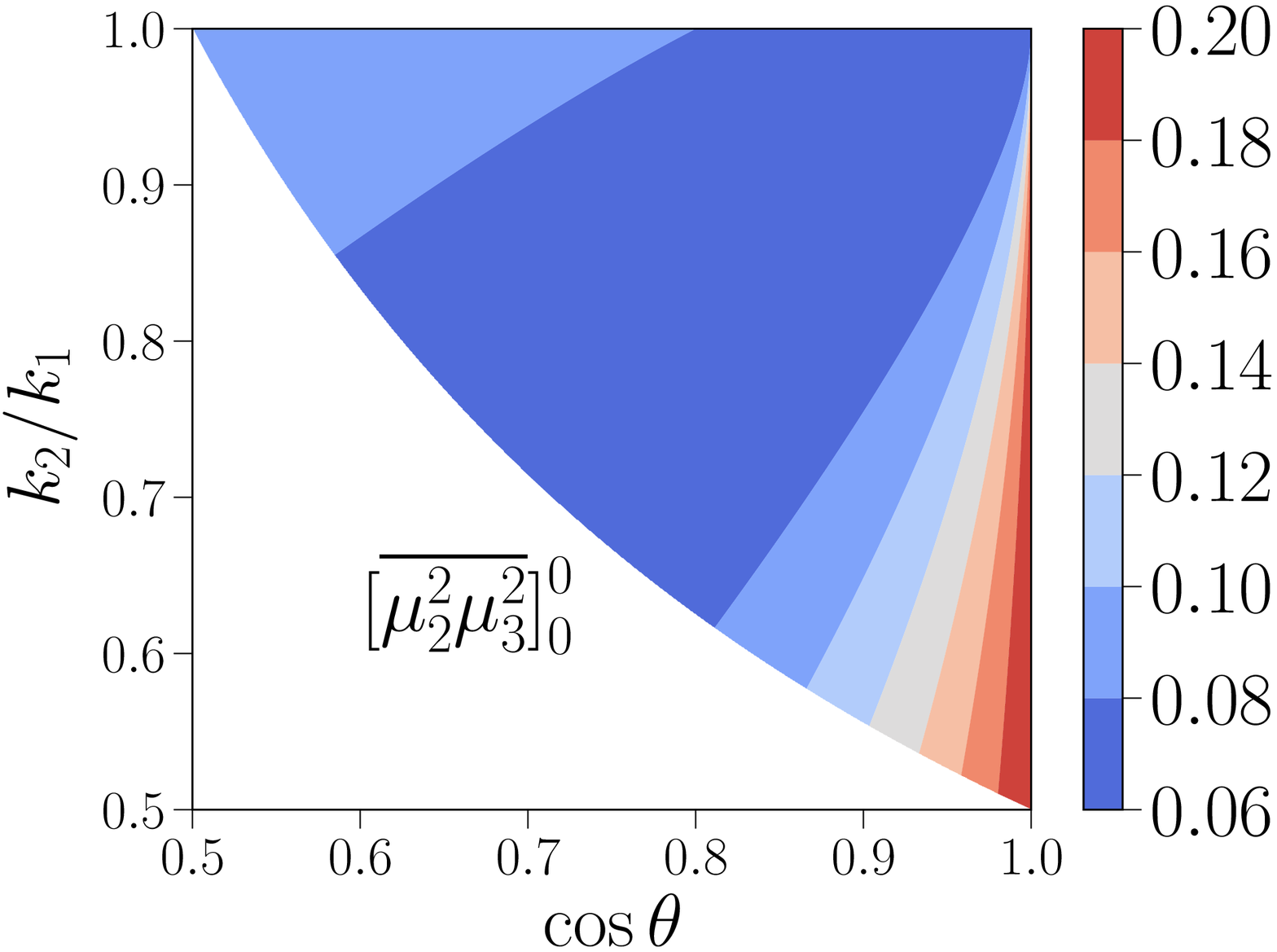}
   \includegraphics[width=0.24\textwidth,angle=0]{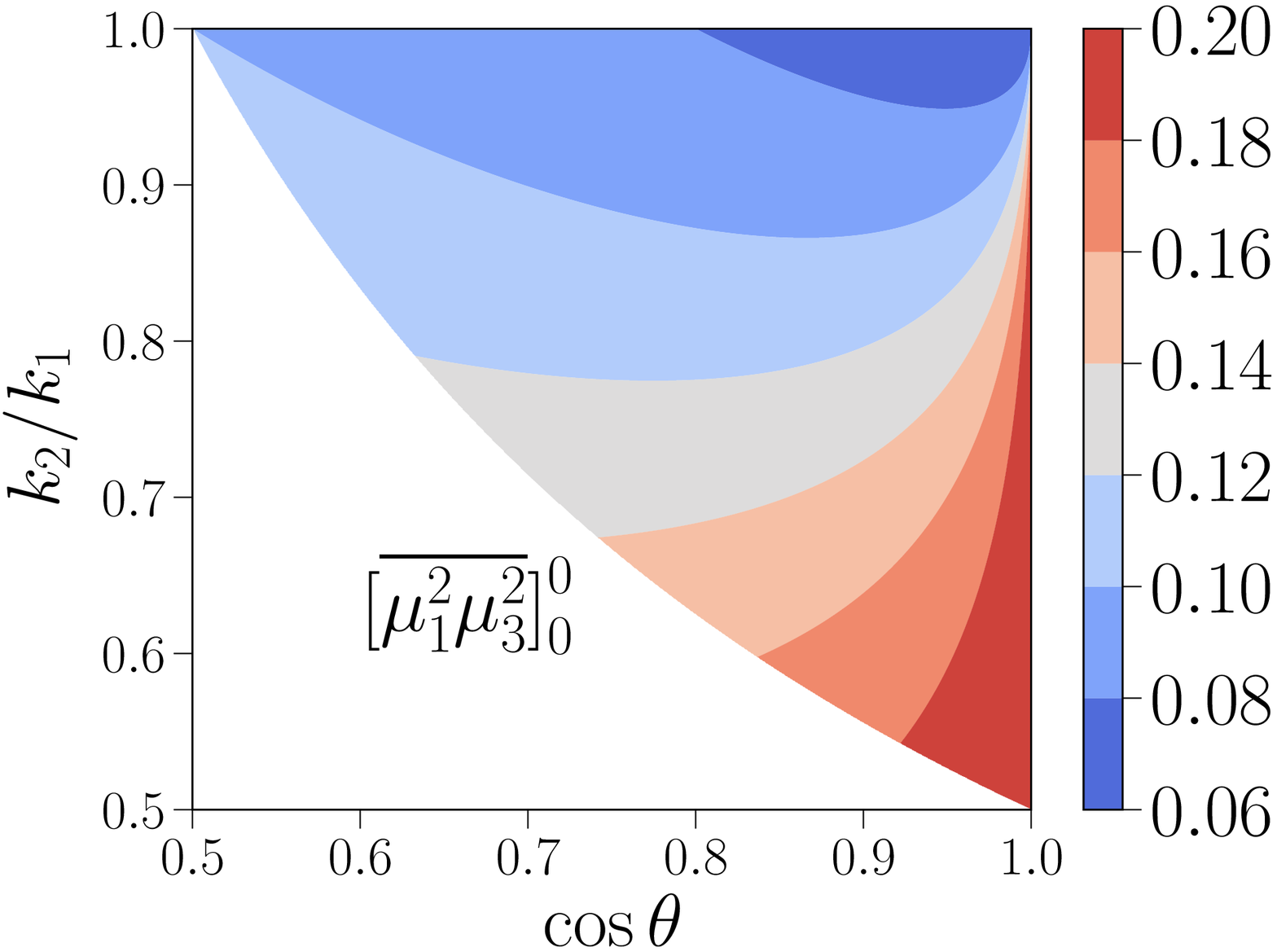}
   \includegraphics[width=0.24\textwidth,angle=0]{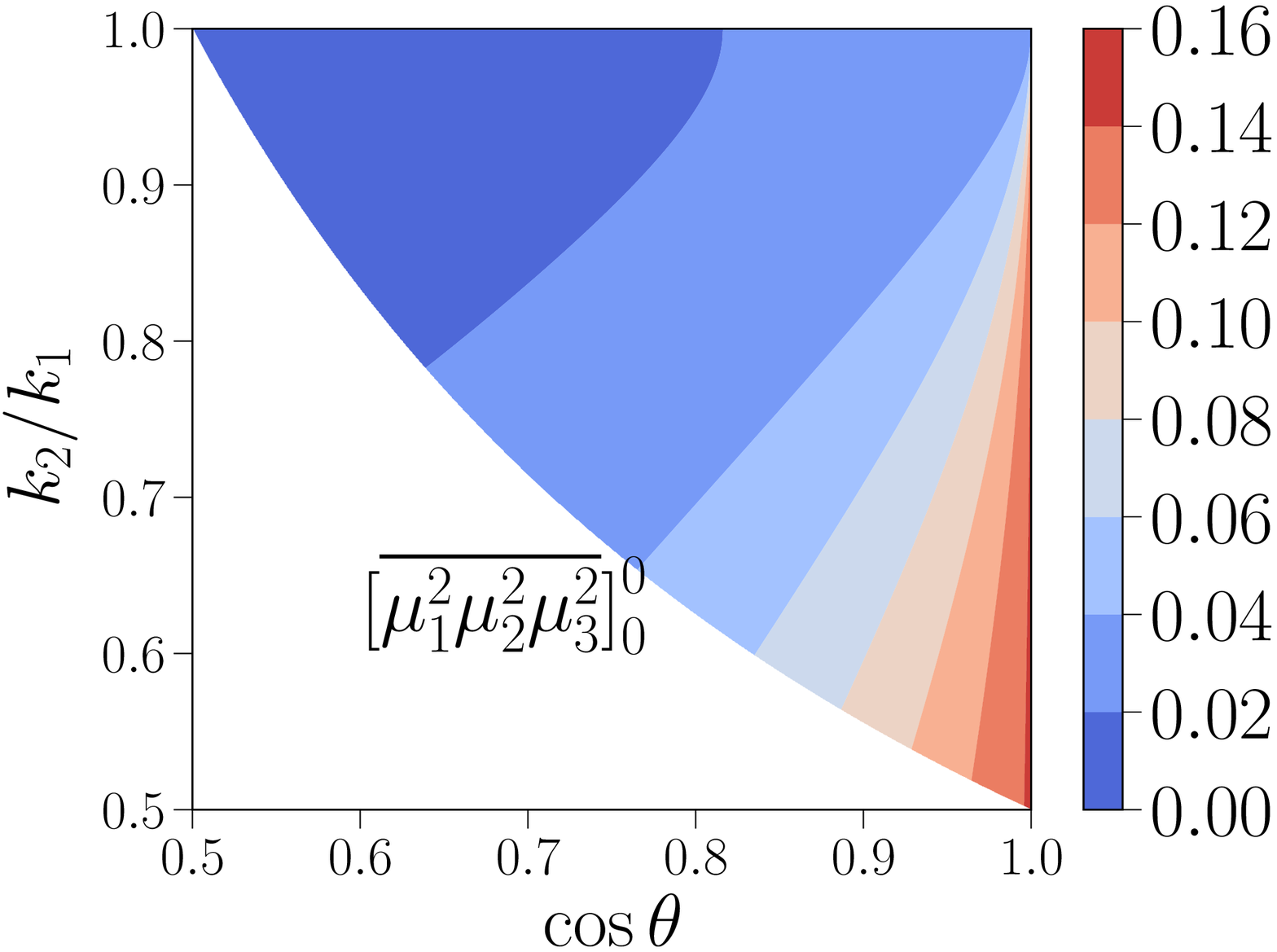}
  
  \caption{These panels show the value of coefficients $[\overline{\mu_1^2 \mu_2^2}]_0^0$, $[\overline{\mu_2^2 \mu_3^2}]_0^0$, $[\overline{\mu_3^2 \mu_1^2}]_0^0$ and $[\overline{\mu_1^2 \mu_2^2 \mu_3^2}]_0^0$ in the equation \eqref{eq:B_qlin1} for $m= 0,\, \ell=0$.}
  \label{fig:monopole}
\end{figure*}

The panels in Figure \ref{fig:monopole} show the coefficients of different components of the quasi-linear model for redshift space bispectra defined by Equation \eqref{eq:B_qlin1}. The four panels of Figure \ref{fig:monopole} show the four coefficients defined in Equations \eqref{eq:B_mu4_1} - \eqref{eq:B_mu6} for the monopole moment of the bispectra i.e. for $m =0$ and $\ell = 0$ in Equation \eqref{eq:B_qlin1}.  

\begin{figure*}
  \includegraphics[width=1\textwidth,angle=0]{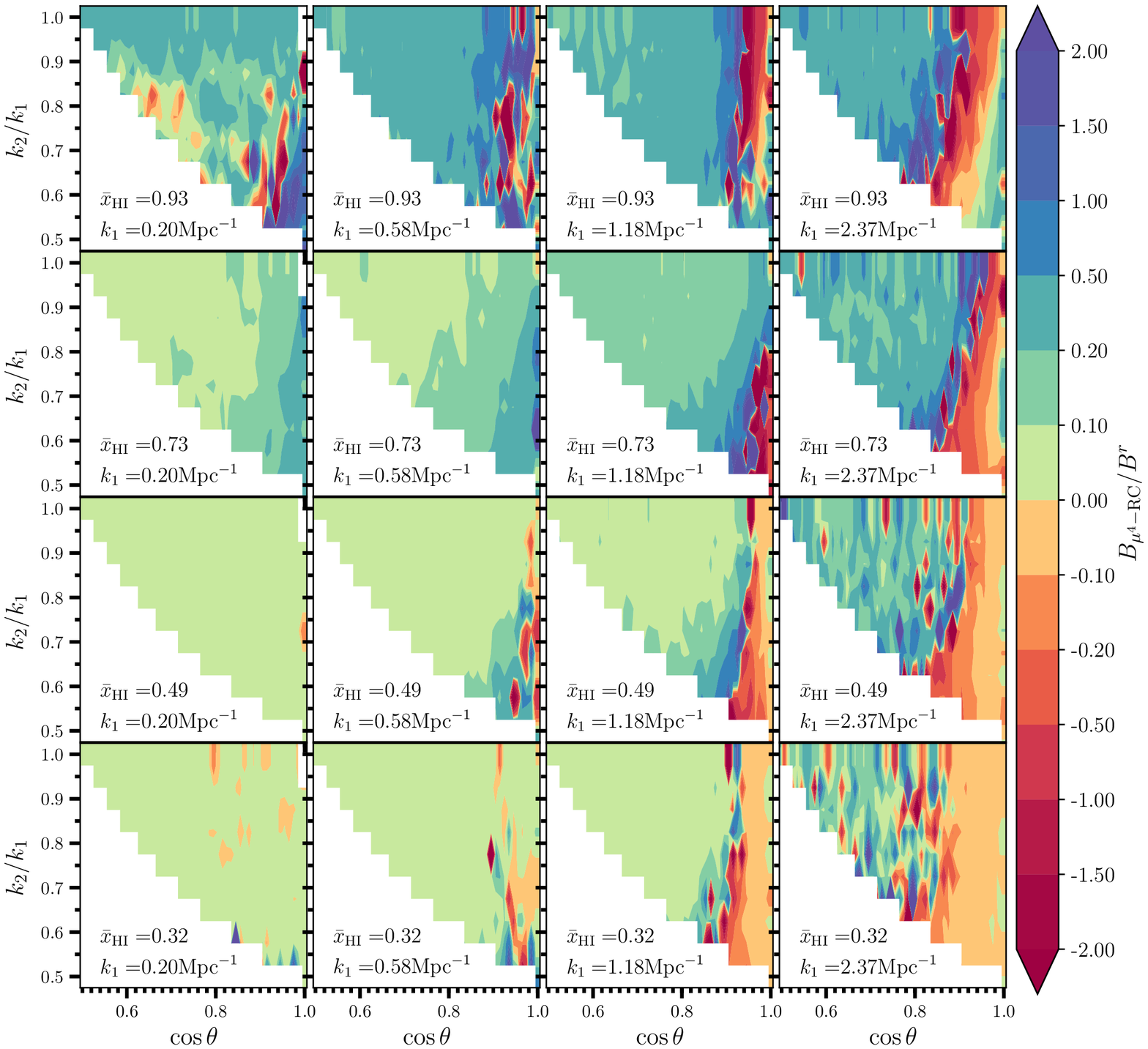}
  \caption{The ratio between ${B_{{\mu^4}-{\rm RC}}}$ and $B^r$  for all unique triangle configurations at four different stages of the EoR and for four different $k_1$ modes.}
  \label{fig:B4_to_Br}
\end{figure*}

\begin{figure*}
  \includegraphics[width=1\textwidth,angle=0]{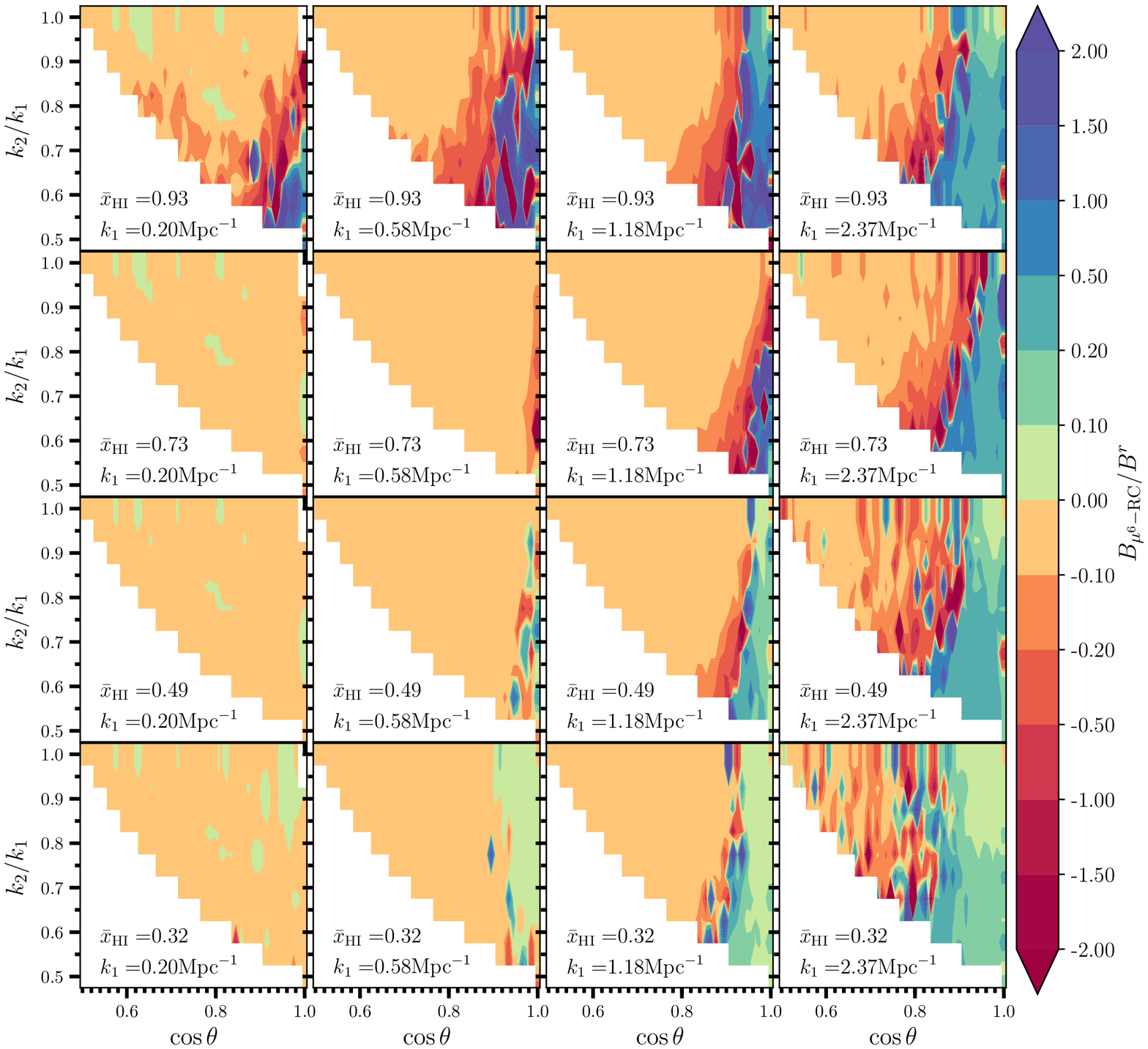}
  \caption{The ratio between the ${B_{{\mu^6}-{\rm RC}}}$  and $B^r$ for all unique triangle configurations at four different stages of the EoR and for four different $k_1$ modes.}
  \label{fig:B6_to_Br}
\end{figure*}

Figure \ref{fig:B4_to_Br} shows the ratio between ${B_{{\mu^4}-{\rm RC}}}$ and $B^r$ for all unique triangle configurations at four different stages of the EoR and for four different $k_1$ modes. Figure \ref{fig:B6_to_Br} shows the ratio between ${B_{{\mu^6}-{\rm RC}}}$ and $B^r$ for all unique triangle configurations at four different stages of the EoR and for four different $k_1$ modes. The ${B_{{\mu^4}-{\rm RC}}}$ and ${B_{{\mu^6}-{\rm RC}}}$  are the two mostly minor correction terms which impact the EoR 21-cm signal bispectra in redshift space. These two figures show the relative contributions of the ${B_{{\mu^4}-{\rm RC}}}$ and ${B_{{\mu^6}-{\rm RC}}}$ groups of RC terms in the redshift space bispectra at different stages of the EoR.


\end{document}